\documentclass{cimart}

\usepackage{bm}

\DeclareMathOperator\opd{d}
\DeclareMathOperator\ope{e}
\DeclareMathOperator\opp{p}
\DeclareMathOperator\opt{t}

\title{On modeling and solving the Boltzmann equation}

\authors{Liliane Basso Barichello}

\authorinfo{Federal University of Rio Grande do Sul, Brazil}{lbaric@mat.ufrgs.br}

\abstract{The Boltzmann equation has been a driving force behind significant mathematical research over the years. Its challenging theoretical complexity, combined with a wide variety of current scientific and technological problems that require numerical simulations based on this model, justifies such interest. This work provides a brief overview of studies and advances related to the solution of the linear Boltzmann equation in one- and two-dimensional spatial dimensions. In particular, relevant aspects of the discrete ordinates approximation of the model are highlighted for neutron and photon transport applications, including nuclear safeguards, nuclear reactor shielding problems, and optical tomography. In addition, a short discussion on rarefied gas dynamics problems, which are relevant, for instance, in the studies of micro-electro-mechanical systems, and their connection with the Linearized Boltzmann Equation, is presented. A primary goal of the work is to establish as much as possible the connections between the different phenomena described by the model and the versatility of the analytical methodology, the ADO method, in providing concise and accurate solutions, which are fundamental for numerical simulations.
}

\keywords{Boltzmann Equation, Mathematical Physics, Numerical Analysis, Transport Theory, Deterministic Methods.}
    
\msc{35Q20 (primary); 65N35 (secondary).}

\VOLUME{34}
\YEAR{2026}
\ISSUE{2}
\NUMBER{5}
\DOI{https://doi.org/10.46298/cm.16372}
\licence{CC BY-NC-ND 4.0}

\editinfo{August 20, 2025}{January 11, 2026}{Tiago Macedo, Jaqueline Mesquita, Mariel Saez, and Rafael Potrie}

\acknowledgments{The author thanks to CNPq of Brazil for partial financial support to this and other works. The author also wants to express her gratitude to all the co-authors referenced in this work, whose collaboration was fundamental to making it possible to develop the ADO method. Very special thanks to C. E. Siewert.}

\begin{document}

\section{Introduction}
\label{intro}

The Boltzmann equation (BE) \cite{Boltzmann1872} has been the reason and motivation for relevant mathematical studies and research over the years \cite{Cercig1988,cedric}. Its challenging theoretical complexity, combined with a wide variety of current problems that require numerical simulations based on this model, justifies such interest.
The fundamental model derived in the context of the kinetic theory of gases concerns a nonlinear integro-differential operator. The treatment of the collision integral operator is known to be a significant challenge. For simulation purposes, the Linearized Boltzmann equation (LBE) or kinetic models derived from the original equation are commonly used. On the other hand, when interactions between particles are neglected and the goal is to describe the transport of a particle in a material medium, the balance of the particles results in a linear model. The linear version of the Boltzmann equation is commonly referred to as the Transport Equation (TE) or Radiative Transfer Equation (RTE), for photon applications.

In this work, we report studies and advances related to the solution of the linear Boltzmann equation in one and two spatial dimensions. We  discuss relevant aspects of neutron and photon transport modeling, such as in nuclear reactor shielding problems and optical tomography.   In addition, we include a brief report on kinetic models of the rarefied gas dynamics, which are relevant, for instance, in the study of micro-electro-mechanical systems, and their connection to the linearized Boltzmann equation (LBE).

 The already noted complexity of the Boltzmann models reinforces the development of deterministic methods that can provide high-quality weak-form solutions.
 The discretization of variables is necessary in both purely numerical approaches and analytical techniques.
 In this context, 
the discrete ordinates method is well-known and is used in established
 codes \cite{lewis1984}
 to deal with angular variables. The solution for selected discrete directions of the particles, along with the numerical approximation of the integral operator by a quadrature scheme, transforms the original problem into a system of first-order partial or ordinary differential equations. 

 In the literature, the discrete ordinates method is often referred to as the $S_N$ method \cite{Davison2}. In fact, in this case, in addition to the angular discretization, spatial variables are also usually discretized. The work of Chandrasekhar \cite{chandra} in astrophysics was a pioneering effort in proposing analytical solutions for the discrete ordinates approximation of the RTE equation in one-dimensional geometry, expressed in terms of the spatial variable. 
In more recent years, the Analytical Discrete Ordinates (ADO) Method, 
\cite{ado, rio}, a version of Chandrasekhar's solution,  was proposed. The technique comprises
a simultaneous set of features that made Chandrasekhar's original ideas computationally viable, as well as expanded the applicability of the method to problems in other areas of transport theory, such as neutron transport and rarefied gas dynamics. It is worth mentioning such features:
the use of
an arbitrary numerical quadrature scheme defined in the half-range, not restricted to the Gauss-Legendre quadrature scheme; the separation constants are obtained through the solution of an eigenvalue problem, instead of searching for roots of polynomials; 
the resulting eigenvalue problem is of half the order of the number of directions; finally, 
a scaling introduced to deal with the exponential terms in the solution prevents ``overflow".

As we discuss throughout the text, this methodology has enabled the successful treatment of numerous problems in various areas of particle transport.

 In two-dimensional Cartesian geometry, the discrete ordinates approximation reduces the integro-differential equation to a system of first-order partial differential equations. Numerical integration on the unit sphere becomes an issue.
In the case of spatial variables, the so-called nodal methods \cite{Badru1985}
 reduce the
 complexity of the model, following the idea of finite volume schemes,
 and are known to perform better in coarser meshes. The connection
 between the spatial and angular grids is
 usually made in the literature through sweep schemes \cite{lewis1984}
 that demand high computational time.

In this framework, the ADO formulation, along with nodal schemes 
(ADO-Nodal) method,  was
 proposed for solving the discrete ordinates approximation of the two\--dimen\-sion\-al linear Boltzmann
(transport) equation \cite{lbb-lc-jf2011, Picoloto2015, lbb-pic-rdc2017}.
 The discrete ordinates equations are transversely integrated over a region (node) of the domain,
 yielding one-dimensional equations for average angular fluxes or intensities,
 in $x$ and $y$-directions.
 The ADO method \cite{ado} is then applied to the one-dimensional equations,
 with approximations for the unknowns (transverse leakage terms)
 on the contours of the regions, to derive explicit solutions for the spatial variables.

Due to the analytical characteristic of the ADO-Nodal method, as first proposed in \cite{lbb-lc-jf2011},
 the integration procedure to generate the nodal equations was defined for the whole interval of the
 $x$ and $y$ variables definition, respectively.
 No domain division into nodes was carried out. Later,
 the idea of local solutions was considered for the treatment of heterogeneous media
 problems \cite{lbb-pic-rdc2017}
since different materials constitute the physical domain.

Among the good features of this formulation, which is a spectral technique, one may cite: the solution of the transverse integrated one-dimensional equations, either in $x$ and $y$-directions are explicitly written in terms of the spatial variables along with the fact that the associated eigenvalue problem is of reduced order, to only half of the number of discrete directions. Significant advantages in comparison with the nodal schemes available then. With these features, the methodology applied to a broad class of neutron and radiation transport problems has advantages in providing fast and accurate solutions. Enhanced performance has been noted for coarser meshes if compared with analogous methodologies.

The scheme does not use sweep. Linear systems must be solved to fully establish the general solutions.
A significant number of discrete directions in the simulations is theoretically recognized as an error control requirement between the discrete ordinate solution and the exact solution \cite{Madsen1971}. Furthermore, from a physical perspective, problems in highly anisotropic scattering media, for example, also impose such a condition, which means a choice of high-order quadrature schemes to approximate the integral term of the equation.
 Quadrature order is a factor that directly affects the linear system.
We implemented and analyzed different approaches to solve the linear system associated with the ADO-Nodal formulation. We analyzed the performance of the schemes applied to problems in different areas of particle transport, such as neutrons and radiation, as well as the
influence of using different schemes for representing the discrete 
ordinates \cite{coam}.

In this work, we present a general overview of the ADO methodology, aiming to highlight advances in the area and contributions to this topic, as already mentioned, of such complexity. We highlight, in particular, different aspects and mathematical tools necessary for the derivation and implementation of the formulation, as well as for error analysis.

Finally, we indicate future challenges and work in progress.

The manuscript is organized as follows: the general integro-differential formulation is introduced in Section \ref{sec:rte}, for photons (RTE). In Sections \ref{planar} and \ref{sec:ado}, the simpler one-dimensional models are discussed and the ADO formulation is derived. Two-dimensional geometry problems are presented in Section \ref{ado-two}, where the multidimensional quadrature schemes are commented, and the ADO-Nodal formulation is developed. Next, in Section~\ref{neutrons}, neutron transport is discussed, highlighting the error in the spatial discretization of the two-dimensional transport equation. Problems in the general area of rarefied gas dynamics, its connection with the LBE, and the application of the ADO method in this area are analyzed in Section \ref{sec:lbe}. A short report on Inverse problems (Section \ref{sec:inverse}) in the field and the final remarks conclude the
text.

\section{The Integro-differential Formulation}
\label{sec:rte}

\begin{figure}[!ht]
\centering
\includegraphics[scale=0.3]{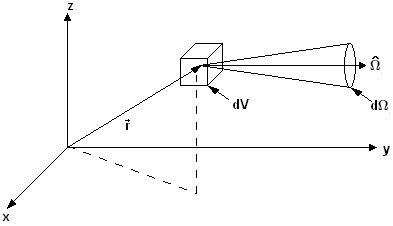}
\hspace{0.8cm}
\includegraphics[scale=0.2]{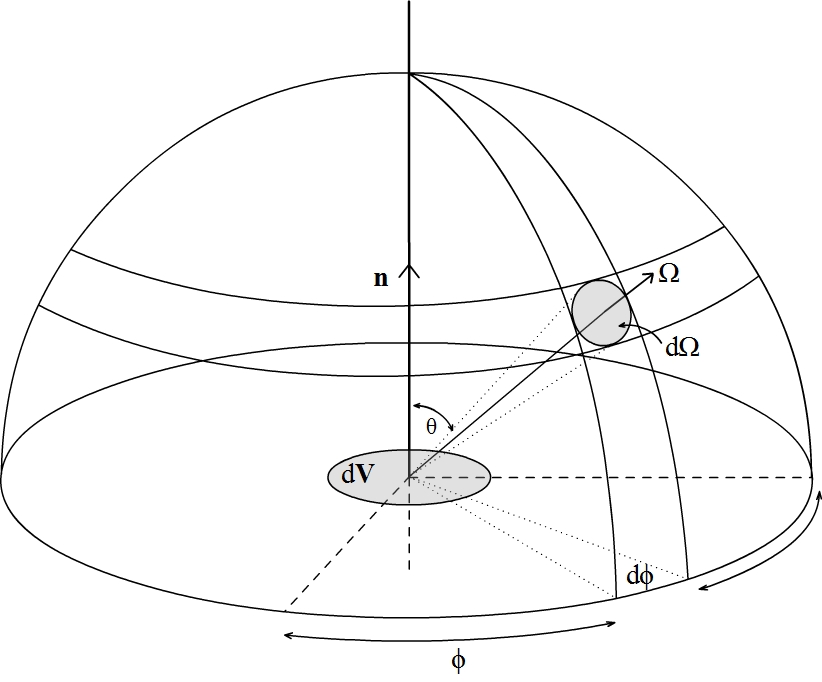}
\caption{Coordinates in the phase space}
\label{balance}
\end{figure}

We begin with the time-independent (RTE),  
  for homogeneous media, which results from a balance of gains and losses in the phase space,
\begin{eqnarray}
\underbrace{\Omega\cdot \nabla I({\bf r},{\boldsymbol\Omega})+\beta I({\bf r},{\boldsymbol \Omega})}_{\mbox{losses}}= \frac{\sigma_s}{4\pi}
\underbrace{\underbrace{\int_{S}p({\boldsymbol\Omega'} \cdot {\boldsymbol\Omega})I({\bf r},{\boldsymbol\Omega'}) \,d{\boldsymbol\Omega'}}_{\mbox{scattering source}} + Q({\bf r},{\boldsymbol \Omega}).}_{\mbox{gains}}\label{eq1}
\end{eqnarray}
 Here $I({\bf r},{\boldsymbol \Omega})$ is the intensity of the radiation in ${\bf r}$ along the direction ${\boldsymbol \Omega}=(\mu,\eta,\xi)$ of the particle as a vector in the unit sphere $S$, with $\mu^2 + \eta^2 + \xi^2 = 1$,
 and (see Fig.~\ref{balance})
 \begin{align}
\mu = \cos\phi \sin\theta, \nonumber\\
\eta = \sin\phi \sin\theta, \nonumber \\
\xi = \cos\theta .
\end{align}

Furthermore, in Eq.~(\ref{eq1}), $\beta=\kappa + \sigma_s$ is the extinction coefficient (absorption + scattering coefficients); 
$p({\boldsymbol\Omega'} \cdot {\boldsymbol\Omega})$ is the scattering phase function;
$Q({\bf r},{\boldsymbol \Omega})$ is an external source term.

In the transport equation/RTE the angular scattering pattern is mathematically described by an expression known as a scattering law/phase function, respectively.
Although for some applications an exact formulation of such a function is known, the representation through a truncated expansion in terms of Legendre polynomials is mostly used in the literature, i.e.,
\begin{equation}
p (\boldsymbol{\Omega}'\cdot \boldsymbol{\Omega}) = p (\cos {\Theta}) = 
\sum_{l=0}^{L} C_{l} P_{l} (\cos{{\Theta}}),
\label{expansao}
\end{equation}
where $\Theta$, the phase-angle, is the angle between incident and scattered directions; $C_{l}$ are the expansion coefficients;  $L$ refers to the degree of anisotropy of the medium; $P_{l}$ are the Legendre polynomial of $l-th$ degree. Using the 
Addition theorem for the Legendre Polynomials \cite{Abramowitz-Stegun1965}
\begin{equation}
p ({\boldsymbol{\Omega}'}\cdot{\boldsymbol{\Omega}})= p (cos\Theta) = \sum_{l=0}^{L} \cdot \sum_{k=0}^{l} (2 -\delta_{0,k}) C_{l} \frac{(l-k)!}{(l+k)!} \cdot P_{l}^{k} (\xi ) \cdot P_{l}^{k} (\xi ') \cdot cos [k (\varphi - \varphi')]. 
\label{eq_harmonicos_cap2}
\end{equation}
Here $\delta_{0,p}$ is the Kronecker's delta and $P_{l}^{p}$ are the associated  Legendre functions of $k$-order and $l$-degree 
\begin{equation}
    P_{l}^{k} (\xi ) =(-1)^{k}(1-\xi^{2})^{\frac{k}{2}} \frac{d^{k}}{d\xi^{k}}(P_{l}(\xi)).
\end{equation}
The Henyey-Greenstein function is widely used to model scattering in  biological tissues, as in optical tomography applications. In this case,
$C_{l}=(2l+1)g^{l}$, being $g$ the asymmetry factor, $g\in (-1,1)$. 
That way $g=0$ indicates isotropic scattering while $0<g<1$ indicates forward scattering and $-1<g<0$ indicates backward scattering.  On the other hand, the exact representation of the Henyey-Greenstein function is of the form
\begin{equation}
    p_{HG}(\Theta)= \frac{1-g^{2}}{\left[1 +g^{2} -2g \cos{\Theta} \right]^{\frac{3}{2}}}.
    \label{HG_exata}
\end{equation}
To complete the formulation, Eq.~(\ref{eq1}) is supplemented with boundary conditions. In this work, we introduce them later in the text.

\section{A ``Simplest'' Problem}
\label{planar}

\begin{figure}[!ht]
\begin{center}
\includegraphics[scale=0.15]{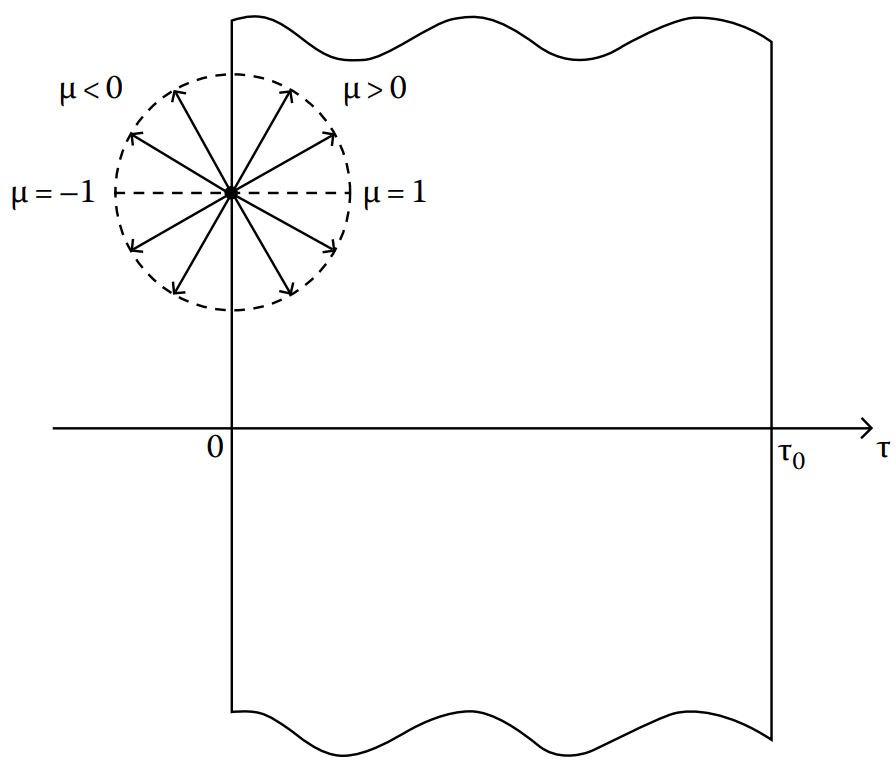}
\caption{The slab geometry}
\label{unid}
\end{center}
\end{figure}

For an introduction to transport theory, it may be helpful to start with the simplest possible geometry: a one-dimensional homogeneous medium with isotropic scattering. The general coordinate system described in Fig.~\ref{balance} is now restricted to what we see in Fig.~\ref{unid}. Under such conditions, we consider the problem given by
\begin{equation}
\mu \frac{\partial}{\partial r} I (r, \mu)  +  I(r, \mu) =
\frac{\varpi}{2} \int_{-1}^{1} I(r, \mu') {\text d} \mu',
\label{eq:slab2}
\end{equation}
\begin{equation}
I(0, \mu) =  F_1(\mu) \quad \mu \in (0,1),
\label{eq:bc2a}
\end{equation}
\begin{equation}
I(\tau_0, -\mu) = F_2(\mu), \quad \mu \in (0,1).
\label{eq:bc2b}
\end{equation}

Here $\tau\in\left(\tau_{a},\tau_{b}\right)$, $\tau$ is the optical thickness, $\mu\in\left[-1,1\right]$ is the cosine of the polar angle as measured from the positive $\tau$-axis, and $\varpi$ is the albedo for single scattering. We assume that the incoming intensities, $F_1(\mu)$ and
$F_2(\mu)$ are known on the boundaries. 
Case's work \cite{case} was a landmark in transport theory, presenting a rigorous mathematical development for addressing such a problem. This procedure is usually called Case's method or the singular eigenfunction expansion method. Known as the only exact solution to this problem, despite some additional results \cite{zweifel} in the '60s and '70s, mainly, it is only viable in idealized cases. However, this solution has been the basis for the development of numerical and semi-analytical methods. More importantly, it emphasizes the difficulty of solving problems of this type, characterized by singular eigenfunctions, associated with continuous and discrete spectrum, and the need for solutions of singular integral equations in assessing orthogonality conditions. 

In other words, to address the fundamental problems in various areas of nuclear science, radiative transfer, and rarefied gas dynamics, alternative approaches needed to be developed.
Two approaches stand out: probabilistic, such as the Monte Carlo method \cite{Haghighat2015}, and deterministic, which includes the discrete ordinates method.

In the next section, we present the development of the ADO formulation for a still basic problem, with the same geometry as in the previous case,  where we introduce complexities such as the treatment of anisotropic media and boundary conditions that include reflection. The idea is to show how much simpler (and powerful at the same time) it is such an approach, in comparison with the Case's solution.

\section{The ADO Solution for One-dimensional Geometry Problems}
\label{sec:ado}

In this work, we apply the Analytical Discrete Ordinates (ADO) Method  \cite{ado} to solve the slab problem defined by Eqs. (\ref{eq:slab2}) to (\ref{eq:bc2b}). Since the ADO formulation was derived in detail for similar models in earlier papers \cite{rio, metti, cht21}, we  briefly develop the main steps here. We begin with the equation valid for anisotropic scattering media,
\begin{equation}\label{rte_isr}
\mu\frac{\partial}{\partial\tau}I\left(\tau,\mu\right)+I\left(\tau,\mu\right)= \frac{\varpi}{2}\sum_{l=0}^{L}\beta_{l}P_{l}\left(\mu\right)\int_{-1}^{1}P_{l}\left(\mu'\right)I\left(\tau,\mu'\right)d\mu'+Q\left(\tau,\mu\right)
\end{equation}
for $\tau\in\left(\tau_{a},\tau_{b}\right)$, $\tau$ is the optical thickness, $\mu\in\left[-1,1\right]$ is the cosine of the polar angle as measured from the positive $\tau$-axis, $\varpi \in [0,1)$ is the albedo for single scattering, 
 $\beta_{l}$ are the coefficients in the $L$th-order expansion of the scattering law and $Q(\tau, \mu)$ is an inhomogeneous source term. We assume boundary conditions defined as
\begin{equation}\label{b0_isr}
I\left(\tau_a,\mu\right) = F_{1}\left(\mu\right)
+\rho_{1}^{s}I\left(\tau_{a},-\mu\right)+2\rho_{1}^{d}\int_{0}^{1}I\left(\tau_a,-\mu'\right)\mu'd\mu'
\end{equation}
and
\begin{equation}\label{bd_isr}
I\left(\tau_{b},-\mu\right) = F_{2}\left(\mu\right)+
\rho_{2}^{s}I\left(\tau_{b},\mu\right)+2\rho_{2}^{d}
\int_{0}^{1}I\left(\tau_{b},\mu'\right)\mu'd\mu',
\end{equation}
for $\mu\in\left(0,1\right]$, which allows for specular and diffuse reflection, with
specular and diffuse reflection coefficients, respectively,
 $\rho_{i}^{s}$ and  $\rho_{i}^{d}$, $i=1,2$.
We also assume that there are known incident distributions at the boundaries $\tau=\tau_a$ and $\tau=\tau_{b}$, given by $F_{1}\left(\mu\right)$ and $F_{2}\left(\mu\right)$, respectively.

We will express the general solution of Eq.~(\ref{rte_isr}) as the sum of the homogeneous plus particular solutions.
Following Barichello \cite{metti}, for developing the homogeneous solution, we assume
 $Q(\tau, \mu)=0$ and rewrite the integral term in equation \eqref{rte_isr}
 as
\begin{equation}\label{ado_int}
\mu\frac{\partial}{\partial\tau}I\left(\tau,\mu\right)+I\left(\tau,\mu\right) = \frac{\varpi}{2}\sum_{l=0}^{L}\beta_{l}P_{l}\left(\mu\right) \int_{0}^{1}P_{l}\left(\mu'\right)\left[I\left(\tau,\mu'\right)+\left(-1\right)^{l})I\left(\tau,-\mu\right)\right]d\mu'.
\end{equation}
Then, we obtain the discrete ordinates version of equation \eqref{ado_int}

\begin{multline}\label{ado_disc}
\pm\mu_{i}\frac{d}{d\tau}\Phi\left(\tau,\pm\mu_{i}\right)+\Phi\left(\tau,\pm\mu_{i}\right)\\ = \frac{\varpi}{2}\sum_{l=0}^{L}\beta_{l}P_{l}\left(\mu_{i}\right) \sum_{k=1}^{N}w_{k}P_{l}\left(\mu_{k}\right)\left[\Phi\left(\tau,\pm\mu_{k}\right)+\left(-1\right)^{l}\Phi\left(\tau,\mp\mu_{k}\right)\right],
\end{multline}
for $i=1,2,\dots,N$, where $\mu_{k}$ are the nodes and $w_{k}$ are the weights of an arbitrary quadrature scheme in the half-range $\left(0,1\right]$.

Equations \eqref{ado_disc} represent a first-order ordinary differential system of equations. Thus, we seek exponential solutions of the form
\begin{equation}\label{ado_exp}
I\left(\tau,\pm\mu_{i}\right)=\phi\left(\nu,\pm\mu_{i}\right)e^{-\tau/\nu}.
\end{equation}
Substituting \eqref{ado_exp} into equations \eqref{ado_disc}, for $i=1,2,\dots,N$, we obtain
\begin{equation}
\left(1\mp\frac{\mu_{i}}{\nu}\right)\phi\left(\nu,\pm\mu_{i}\right)= \frac{\varpi}{2}\sum_{l=0}^{L}\beta_{l}P_{l}\left(\mu_{i}\right) 
\sum_{k=1}^{N}w_{k}P_{l}\left(\mu_{k}\right)\left[\phi\left(\nu,\pm\mu_{i}\right)+\left(-1\right)^{l}\phi\left(\nu,\mp\mu_{i}\right)\right],
\end{equation}
which can be expressed in matrix notation as

\begin{equation}\label{ado_matrix}
\left(\mathbf{I}_{N}\mp\frac{1}{\nu}\mathbf{M}\right)\mathbf{\Phi}_{\pm}\left(\nu_{r}\right)= \frac{\varpi}{2}\sum_{l=0}^{L}\beta_{l}\mathbf{\Pi}\left(l\right)\mathbf{\Pi}\left(l\right)^{T}\mathbf{W}\left[\mathbf{\Phi}_{\pm}\left(\nu\right)+\left(-1\right)^{l}\mathbf{\Phi}_{\mp}\left(\nu\right)\right],
\end{equation}
where

\begin{equation}
\mathbf{\Phi}_{\pm}\left(\nu\right)=\left[\phi\left(\nu,\pm\mu_{1}\right)\quad\phi\left(\nu,\pm\mu_{2}\right)\quad\cdots\quad \phi\left(\nu\pm\mu_{N}\right)\right]^{T}
\end{equation}
and

\begin{equation}
\mathbf{\Pi}\left(l\right)=\left[P_{l}\left(\mu_{1}\right)\quad P_{l}\left(\mu_{2}\right)\quad \cdots\quad P_{l}\left(\mu_{N}\right)\right]^{T}
\end{equation}
are $N\times1$ vectors. In addition,
\begin{equation}
\mathbf{M}=\text{diag}\left\{\mu_{1},\mu_{2},\cdots,\mu_{N}\right\}
\end{equation}
and
\begin{equation}
\mathbf{W}=\text{diag}\left\{w_{1},w_{2},\cdots,w_{N}\right\}
\end{equation}
are $N\times N$ diagonal matrices. Here, we have used $\mathbf{I}_{N}$ to denote the $N\times N$ identity matrix and $T$ to denote the transpose operation.

Now, we define the $N\times1$ vectors

\begin{equation}\label{ado_u}
\mathbf{U}\left(\nu\right)=\mathbf{\Phi}_{+}\left(\nu\right)+\mathbf{\Phi}_{-}\left(\nu\right)
\end{equation}
and
\begin{equation}\label{ado_v}
\mathbf{V}\left(\nu\right)=\mathbf{\Phi}_{+}\left(\nu\right)-\mathbf{\Phi}_{-}\left(\nu\right),
\end{equation}
and we add the two sets of equations given in \eqref{ado_matrix}, to find that

\begin{multline}
\left(\mathbf{I}_{N}-\frac{1}{\nu}\mathbf{M}\right)\mathbf{\Phi}_{+}\left(\nu\right)+\left(\mathbf{I}_{N}+\frac{1}{\nu}\mathbf{M}\right)\mathbf{\Phi}_{-}\left(\nu\right)=  \\ \frac{\varpi}{2}\sum_{l=0}^{L}\beta_{l}\mathbf{\Pi}\left(l\right)\mathbf{\Pi}\left(l\right)^{T}\mathbf{W}\left[\mathbf{\Phi}_{+}\left(\nu\right)+\left(-1\right)^{l}\mathbf{\Phi}_{-}\left(\nu\right)+\mathbf{\Phi}_{-}\left(\nu\right)+\left(-1\right)^{l}\mathbf{\Phi}_{+}\left(\nu\right)\right],
\end{multline}
which can be rewritten as

\begin{equation}\label{ado_eig1}
\left(\mathbf{I}_{N}-\frac{\varpi}{2}\sum_{l=0}^{L}\beta_{l}\mathbf{\Pi}\left(l\right)\mathbf{\Pi}\left(l\right)^{T}\mathbf{W}\left[1+\left(-1\right)^{l}\right]\right)\mathbf{M}^{-1}\mathbf{MU}\left(\nu\right)=\frac{1}{\nu_{r}}\mathbf{MV}\left(\nu\right) .
\end{equation}

Since $\mu_{i}>0$ for $i=1,2,\dots,N$, we note that the inverse of the matrix $\mathbf{M}$, denoted by $\mathbf{M}^{-1}$, is well defined.
Analogously, we subtract the two sets of equations in \eqref{ado_matrix},
to obtain
\begin{equation}
\mathbf{V}\left(\nu\right)-\frac{1}{\nu}\mathbf{MU}\left(\nu\right) = \frac{\varpi}{2}\sum_{l=0}^{L}\beta_{l}\mathbf{\Pi}\left(l\right)\mathbf{\Pi}\left(l\right)^{T}\mathbf{W}\left[1-\left(-1\right)^{l}\right]\mathbf{V}\left(\nu\right),
\end{equation}
which we can rewrite as

\begin{equation}\label{ado_eig2}
\left(\mathbf{I}_{N}-\frac{\varpi}{2}\sum_{l=0}^{L}\beta_{l}\mathbf{\Pi}\left(l\right)\mathbf{\Pi}\left(l\right)^{T}\mathbf{W}\left[1-\left(-1\right)^{l}\right]\right)\mathbf{M}^{-1}\mathbf{MV}\left(\nu\right)=\frac{1}{\nu}\mathbf{MU}\left(\nu\right) .
\end{equation}
If we define the $N\times1$ vectors

\begin{equation}\label{ado_x}
\mathbf{X}\left(\nu\right) = \mathbf{MU}\left(\nu\right),
\end{equation}
\begin{equation}\label{ado_y}
\mathbf{Y}\left(\nu\right) = \mathbf{MV}\left(\nu\right)
\end{equation}
and the $N\times N$ matrices
\begin{equation}
\mathbf{A} = \left(\mathbf{I}_{N}-\frac{\varpi}{2}\sum_{l=0}^{L}\beta_{l}\mathbf{\Pi}\left(l\right)\mathbf{\Pi}\left(l\right)^{T}\mathbf{W}\left[1+\left(-1\right)^{l}\right]\right)\mathbf{M}^{-1}
\end{equation}
and
\begin{equation}
\mathbf{B} = \left(\mathbf{I}_{N}-\frac{\varpi}{2}\sum_{l=0}^{L}\beta_{l}\mathbf{\Pi}\left(l\right)\mathbf{\Pi}\left(l\right)^{T}\mathbf{W}\left[1-\left(-1\right)^{l}\right]\right)\mathbf{M}^{-1},
\end{equation}
we can express Eqs.~\eqref{ado_eig1} and \eqref{ado_eig2} in the form

\begin{equation}\label{ado_eig3}
\mathbf{A}\mathbf{X}\left(\nu\right)=\frac{1}{\nu}\mathbf{Y}\left(\nu\right)
\end{equation}
and
\begin{equation}\label{ado_eig4}
\mathbf{B}\mathbf{Y}\left(\nu\right)=\frac{1}{\nu}\mathbf{X}\left(\nu\right),
\end{equation}
respectively. From the last two expressions, we obtain two eigenvalue problems

\begin{equation}\label{ado_eigen}
\left(\mathbf{B}\mathbf{A}\right)\mathbf{X}\left(\nu\right)=\frac{1}{\nu^{2}}\mathbf{X}\left(\nu_{r}\right),
\end{equation}

\begin{equation}\label{ado_eigen2}
\left(\mathbf{A}\mathbf{B}\right)\mathbf{Y}\left(\nu\right)=\frac{1}{\nu^{2}}\mathbf{Y}\left(\nu_{r}\right).
\end{equation}
We note that the separation constants, $\nu$, will appear in pairs $\left(\pm\right)$.  In addition, the eigenvalue problems \eqref{ado_eigen} and
 \eqref{ado_eigen2} are of half-order in comparison with the ones obtained in the standard discrete ordinates methods, which are based on full-range quadrature schemes.
 Continuing, from equation \eqref{ado_eig3} we have
\begin{equation}
\nu \mathbf{A}\mathbf{X}\left(\nu\right)=\mathbf{Y}\left(\nu\right).
\end{equation}
If we add  $\mathbf{X}\left(\nu\right)$ to both sides of this expression
 and use Eqs. \eqref{ado_x} and \eqref{ado_y}, we find that

%\begin{equation}
%\left(\mathbf{I}_{N}+\nu \mathbf{A}\right)\mathbf{X}\left(\nu\right)=\mathbf{X}\left(\nu\right)+\mathbf{Y}\left(\nu\right),
%\end{equation}

%and, using equations \eqref{ado_x} and \eqref{ado_y}, we find that

\begin{equation}
\left(\mathbf{I}_{N}+\nu \mathbf{A}\right)\mathbf{X}\left(\nu\right)=\mathbf{M}\left[\mathbf{U}\left(\nu\right)+\mathbf{V}\left(\nu\right)\right].
\end{equation}
From Eqs. \eqref{ado_u} and \eqref{ado_v}, we rewrite the above equation as

\begin{equation}
\left(\mathbf{I}_{N}+\nu \mathbf{A}\right)\mathbf{X}\left(\nu\right)=2\mathbf{M}\mathbf{\Phi}_{+}\left(\nu\right).
\end{equation}
Thus, we obtain the eigenfunctions 
\begin{equation}\label{ado_phi+}
\mathbf{\Phi}_{+}\left(\nu\right)=\frac{1}{2}\mathbf{M}^{-1}\left(\mathbf{I}_{N}+\nu_{r} \mathbf{A}\right)\mathbf{X}\left(\nu\right).
\end{equation}

Analogously, subtracting $\mathbf{X}\left(\nu\right)$ in Eq. \eqref{ado_eig3}
 and using Eqs. \eqref{ado_x} and \eqref{ado_y} we obtain
 \begin{equation}\label{ado_phi-}
\mathbf{\Phi}_{-}\left(\nu\right)=\frac{1}{2}\mathbf{M}^{-1}\left(\mathbf{I}_{N}-\nu \mathbf{A}\right)\mathbf{X}\left(\nu\right).
\end{equation}
Now, if we let

\begin{equation}
\mathbf{I}_{\pm}^{h}\left(\tau\right) = \left[{I}^{h}\left(\tau,\pm\mu_{1}\right)\quad {I}^{h}\left(\tau,\pm\mu_{2}\right)\quad\cdots\quad {I}^{h}\left(\tau,\pm\mu_{N}\right)\right]^{T},
\end{equation}
from the solution of the eigenvalue problem, along with the elementary solutions, \eqref{ado_phi+} and \eqref{ado_phi-}, we can write the discrete ordinates solution for the homogeneous version of Eq.~\eqref{rte_isr} in the vector form
\begin{equation}\label{ado_sol}
\mathbf{I^{h}}\left(\tau\right)=\sum_{j=1}^{N} \left[A_{j}\mathbf{\Phi}_{\pm}\left(\nu_{j}\right)e^{-\left(\tau-\tau_{a}\right)/\nu_{j}}+B_{j}\mathbf{\Phi}_{\mp}\left(\nu_{j}\right)e^{-\left(\tau_{b}-\tau\right)/\nu_{j}}\right],
\end{equation}
where $A_{j}$ and $B_{j}$ are the arbitrary coefficients to be determined next. We note that, in writing the solution in this way, we avoid the computational issue of
exponential terms which can lead to overflow \cite{rio,metti}.

To complete our solution, concerning to the particular solutions we follow \cite{lbb2000}. We omit here the details of the derivation, which was based on the infinite-medium Green's function.
 We express the components of the $N\times 1$ vector
\begin{equation}
\mathbf{I}_{\pm}^{p}\left(\tau\right) = \left[I^{p}\left(\tau,\pm\mu_{1}\right)\quad I^{p}\left(\tau,\pm\mu_{2}\right)\quad\cdots\quad I^{p}\left(\tau,\pm\mu_{N}\right)\right]^{T},
\end{equation}
in the form
\begin{equation}\label{part_sol}
\mathbf{I}_{\pm}^{p}\left(\tau\right)=\sum_{j=1}^{N}A_{j}\left(\tau\right)\mathbf{\Phi}_{\pm}\left(\nu_{j}\right)+B_{j}\left(\tau\right)\mathbf{\Phi}_{\mp}\left(\nu_{j}\right),
\end{equation}
where
\begin{equation}\label{cj1}
A_{j}\left(\tau\right)=\frac{1}{{N}\left(\nu_{j}\right)}\sum_{k=1}^{N}w_{k} \int_{\tau_{a}}^{\tau}\left[Q\left(\tau',\mu_{k}\right)\phi\left(\nu_{j},\mu_{k}\right)+Q\left(\tau',-\mu_{k}\right)\phi\left(\nu_{j},-\mu_{k}\right)\right]e^{-\left(\tau-\tau'\right)/\nu_{j}}d\tau'
\end{equation}
and

\begin{equation}\label{dj1}
B_{j}\left(\tau\right)=\frac{1}{N\left(\nu_{j}\right)}\sum_{k=1}^{N}w_{k} \int_{\tau}^{\tau_{b}}\left[Q\left(\tau',\mu_{k}\right)\phi\left(\nu_{j},-\mu_{k}\right)+Q\left(\tau',-\mu_{k}\right)\phi\left(\nu_{j},\mu_{k}\right)\right]e^{-\left(\tau'-\tau\right)/\nu_{j}}d\tau'.
\end{equation}
Here
\begin{equation}
        {N}\left(\nu_{j}\right)=\sum_{k=1}^{N}w_{k}\mu_{k}\left[\phi\left(\nu_j,\mu_{k}\right)^{2}-\phi\left(\nu_{j},-\mu_{k}\right)^{2}\right].
\end{equation}

At this point, we can represent, in a vector form, the general solution of
 Eq. \eqref{rte_isr},
\begin{equation}\label{ado_csol}
\mathbf{I}_{\pm}\left(\tau\right)=\sum_{j=1}^{N}\left[A_{j}\mathbf{\Phi}_{\pm}\left(\nu_{j}\right)e^{-\left(\tau-\tau_{a}\right)/\nu_{j}}+B_{j}\mathbf{\Phi}_{\mp}\left(\nu_{j}\right)e^{-\left(\tau_{b}-\tau\right)/\nu_{j}}\right]+\mathbf{I}_{\pm}^{p}\left(\tau\right).
\end{equation}

To  completely establish the solution, we need to determine the superposition coefficients, $A_{j}$ and $B_{j}$. These constants can be calculated as the solution of a $2N\times2N$ linear system of equations, given below, obtained from the application of the boundary conditions, Eqs. \eqref{b0_isr} and \eqref{bd_isr},

%\begin{equation}\label{disc_b0}  %\sum_{j=1}^{N}A_{j}\left[\mathbf{\Phi}_{+}\left(\nu_{j}\right)
%$-\bm{\rho}_{1}^{s}\mathbf{\Phi}_{-}\left(\nu_{j}\right)\right]
%+B_{j}\left[\mathbf{\Phi}_{-}\left(\nu_{j}\right)-%\bm{\rho}_{1}^{s}\mathbf{\Phi}_{+}\left(\nu_{j}\right)\right]e^{-%\tau_{a}/\nu_{j}}=
%        \mathbf{F}_{1}-\mathbf{I}_{+}^{p}\left(\tau_a\right),
%\end{equation}

%\begin{multline}\label{disc_bd}
%\sum_{j=1}^{N}A_{j}\left[\mathbf{\Phi}_{-}\left(\nu_{j}\right)-%\bm{\rho}_{2}^{s}\mathbf{\Phi}_{+}\left(\nu_{j}\right)\right]e^{-%\left(\tau_{b}-%\tau_{a}\right)/\nu_{j}}+B_{j}\left[\mathbf{\Phi}_{+}\left(\nu_{j}\right%)-\bm{\rho}_{2}^{s}\mathbf{\Phi}_{-}\left(\nu_{j}\right)\right]= 
%\mathbf{F}_{2}-\mathbf{I}_{-}^{p}\left(\tau_{b}\right).
%\end{multline}

\begin{equation}\label{disc_b0}
\begin{split}%\label{disc_b0}
	\sum_{j=1}^{N}A_{j}\left[\mathbf{\Phi}_{+}\left(\nu_{j}\right)-\bm{\rho}_{1}^{s}\mathbf{\Phi}_{-}\left(\nu_{j}\right)-2\left(\mathbf{\Upsilon}^{T}\mathbf{M}\mathbf{W}\mathbf{\Phi}_{-}\left(\nu_{j}\right)\right)\bm{\rho}_{1}^{d}\mathbf{\Upsilon}\right]&\\+B_{j}\left[\mathbf{\Phi}_{-}\left(\nu_{j}\right)-\bm{\rho}_{1}^{s}\mathbf{\Phi}_{+}\left(\nu_{j}\right)-2\left(\mathbf{\Upsilon}^{T}\mathbf{M}\mathbf{W}\mathbf{\Phi}_{+}\left(\nu_{j}\right)\right)\bm{\rho}_{1}^{d}\mathbf{\Upsilon}\right]&e^{-\tau_{a}/\nu_{j}}\\=
	\mathbf{F}_{1}-\mathbf{I}_{+}^{p}\left(\tau_a\right)+\bm{\rho}_{1}^{s}\mathbf{I}_{-}^{p}\left(\tau_a\right)+&2\left(\mathbf{\Upsilon}^{T}\mathbf{M}\mathbf{W}\mathbf{I}_{-}^{p}\left(\tau_a\right)\right)\bm{\rho}_{1}^{d}\mathbf{\Upsilon}
\end{split}
\end{equation}

and

\begin{equation}\label{disc_bd}
\begin{split}%\label{disc_bd}
\sum_{j=1}^{N}A_{j}\left[\mathbf{\Phi}_{-}\left(\nu_{j}\right)-\bm{\rho}_{2}^{s}\mathbf{\Phi}_{+}\left(\nu_{j}\right)-2\left(\mathbf{\Upsilon}^{T}\mathbf{M}\mathbf{W}\mathbf{\Phi}_{+}\left(\nu_{j}\right)\right)\bm{\rho}_{2}^{d}\mathbf{\Upsilon}\right]&e^{-\left(\tau_{b}-\tau_{a}\right)/\nu_{j}}\\+B_{j}\left[\mathbf{\Phi}_{+}\left(\nu_{j}\right)-\bm{\rho}_{2}^{s}\mathbf{\Phi}_{-}\left(\nu_{j}\right)-2\left(\mathbf{\Upsilon}^{T}\mathbf{M}\mathbf{W}\mathbf{\Phi}_{-}\left(\nu_{j}\right)\right)\bm{\rho}_{2}^{d}\mathbf{\Upsilon}\right]&\\=
\mathbf{F}_{2}-\mathbf{I}_{-}^{p}\left(\tau_{b}\right)+\bm{\rho}_{2}^{s}\mathbf{I}_{+}^{p}\left(\tau_{b}\right)+&2\left(\mathbf{\Upsilon}^{T}\mathbf{M}\mathbf{W}\mathbf{I}_{+}^{p}\left(\tau_{b}\right)\right)\bm{\rho}_{2}^{d}\mathbf{\Upsilon}.
\end{split}
\end{equation}

Here we define, for $i=1,2$ the $N \times 1$ vectors
\begin{equation}
        \mathbf{F}_{i}=\left[F_{i}\left(\mu_{1}\right) \quad F_{i}\left(\mu_{2}\right) \quad \cdots \quad F_{i}\left(\mu_{N}\right)\right]^{T},
\end{equation}
\begin{equation}\label{vet1}
\mathbf{\Upsilon}=\left[1 \quad 1 \quad \cdots \quad 1\right]^{T}
\end{equation}
and the $N\times N$ diagonal matrices
\begin{equation}
\bm{\rho}_{i}^{s}=\text{diag}\{\rho_{i}^{s},\rho_{i}^{s},\cdots,\rho_{i}^{s}\},
\end{equation}
%\begin{equation}
%\bm{\rho}_{2}^{s}=\text{diag}\%{\rho_{2}^{s},\rho_{2}^{s},\cdots,\rho_{2}^{s}\},
%\end{equation}
\begin{equation}
\bm{\rho}_{i}^{d}=\text{diag}\{\rho_{i}^{d},\rho_{i}^{d},\cdots,\rho_{i}^{d}\}.
\end{equation}
%\begin{equation}
%\bm{\rho}_{2}^{d}=\text{diag}\%%{\rho_{2}^{d},\rho_{2}^{d},\cdots,\rho_{2}^{d}\}.
%\end{equation}

Once we solve the linear system to obtain the coefficients,
 we can express the components of the solution vector,
  \eqref{ado_csol},  as
\begin{equation}
I(\tau, \pm \mu) = \sum_{j=1}^{N}\left[A_{j}{\phi}\left(\nu_{j}, \pm \mu\right)e^{-\left(\tau-\tau_{a}\right)/\nu_{j}}+B_{j}{\phi}\left(\nu_{j}, \mp \mu \right)e^{-\left(\tau_{b}-\tau\right)/\nu_{j}}\right] 
 +{I}^{p}\left(\tau, \pm \mu \right).
\end{equation}
Using this ADO solution for the problem given by Eqs. \eqref{rte_isr}, 
 \eqref{b0_isr} and \eqref{bd_isr} we can evaluate the radiation density
\begin{equation}
\rho\left(\tau\right) = \int_{-1}^{1}I\left(\tau,\mu\right)d\mu,
\end{equation}
 as
\begin{equation}
\label{eq:adodensity}
\rho\left(\tau\right) = \int_{0}^{1}[I^{h}\left(\tau,\mu\right)+I^{h}\left(\tau,-\mu\right)] d\mu +\int_{0}^{1}[I^{p}\left(\tau,\mu\right)+I^{p}\left(\tau,-\mu\right)]d\mu,
\end{equation}
for $\tau\in\left[\tau_{a},\tau_{b}\right]$. Then if we approximate the integral term using the defined quadrature scheme in the half-range interval,
 the density can be calculated by
\begin{equation}\label{density}
\begin{split}
\rho\left(\tau\right) =
\sum_{j=1}^{N}&\left[A_{j}e^{-\left(\tau-\tau_{a}\right)/\nu_{j}}+B_{j}e^{-\left(\tau_{b}-\tau\right)/\nu_{j}}+A_{j}\left(\tau\right)+B_{j}\left(\tau\right)\right]\Phi_{0}\left(\nu_{j}\right)
\end{split}
\end{equation}
with
\begin{equation}
\label{eq:fizero}
\Phi_{0}\left(\nu_{j}\right)=\sum_{k=1}^{N}w_{k}\left[\phi\left(\nu_{j},\mu_{k}\right)+\phi\left(\nu_{j},-\mu_{k}\right)\right].
\end{equation}

\subsection{Numerical Results and Contributions}
The ADO formulation for problems in one-dimensional geometry was developed in this section in the context of the RTE. In addition to the works already cited, the methodology has been applied to a wide range of problems in various areas of transport theory by different authors. It is worth listing here the treatment of models which include polarization effects \cite{Barichello1999b} of interest in astrophysics, and that, according to our knowledge, has not been solved before; the challenging searchlight problem \cite{Barichello2000b}; ADO for different geometries \cite{cylinder}; medical applications \cite{cht21}; the key issues of criticality \cite{adocomplex} and safeguards \cite{ane2025} in nuclear engineering; to mention just a few. Later in this work, applications in the field of rarefied gas dynamics will be discussed. We do not list numerical results generated by the solution here. They can be checked in the listed references. However, a central point to note is that in all these applications, the ADO solution proved to be concise and accurate, requiring a lower computational cost compared to what was available in the literature. The natural challenge, considering more realistic problems, was to extend the idea to multidimensional problems, which are the subject of the next section.

\section{The ADO Solution for Two-dimensional Geometry Problems}
\label{ado-two}

Purely numerical methods dominate simulations for multidimensional transport models. 
The use of analytical techniques is limited and very challenging.
However, they are of great importance for generating benchmark solutions. 
One of the first issues regarding discrete ordinate methods, which already differ greatly from one-dimensional problems, is the definition 
of numerical quadrature schemes for integration on the unit sphere. Madsen \cite{Madsen1971} identified the truncation error in the approximation 
of the integral term of the equation as a fundamental point in the convergence of the discrete-ordinate solution.

\subsection{Numerical Quadrature Schemes}
{ \quad }
\label{sec:quadrature}

\begin{figure}[!ht]
\centering
\includegraphics[scale=0.18]{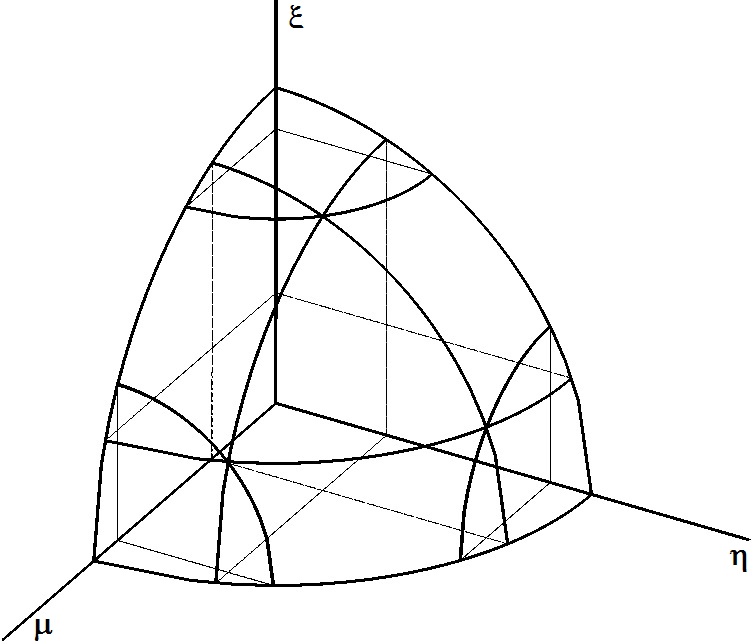}
\includegraphics[scale=0.18]{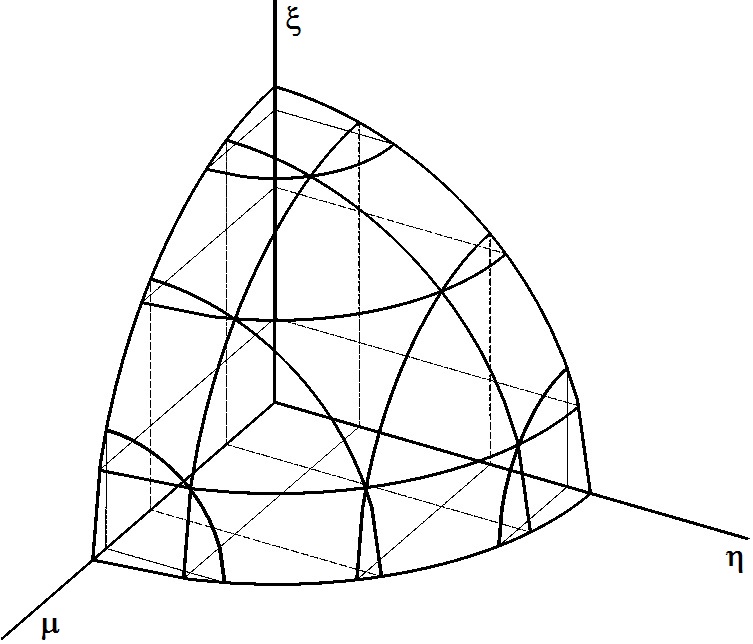}
\caption{Level Symmetric Quadrature Scheme}
\label{levelset}
\end{figure}

The choice of an appropriate  quadrature scheme to approximate the
 integral term in Eq.~(\ref{eq1}) is a relevant issue, in particular for multidimensional problems. 
 Traditionally, the quadratures were required to preserve the reflective and rotational symmetry properties
 and, in this context,  the classical Level Symmetric ($LQ_{N}$) 
quadrature scheme \cite{lewis1984}
 is widely used for dealing with multidimensional discrete ordinates problems (see Fig.~\ref{levelset}). 
 It is well known, however, that such a scheme is limited  to order
 $N \le 20$. The number of directions per octant is given by  $$M=\frac{N(N+2)}{8}.$$

In our studies, we have additionally considered triangular ($P_N T_N S_N$) and 
 quadrangular ($P_N T_N$) Legendre-Chebyshev schemes \cite{pleg-cheb,cacuci} and 
 Quadruple Range ($QR$) schemes \cite{abu1}, which enable the generation of numerical results for much higher quadrature orders, even if
 some (rotational) symmetry properties are relaxed.

\begin{figure}[!ht]
\centering
\includegraphics[scale=0.4]{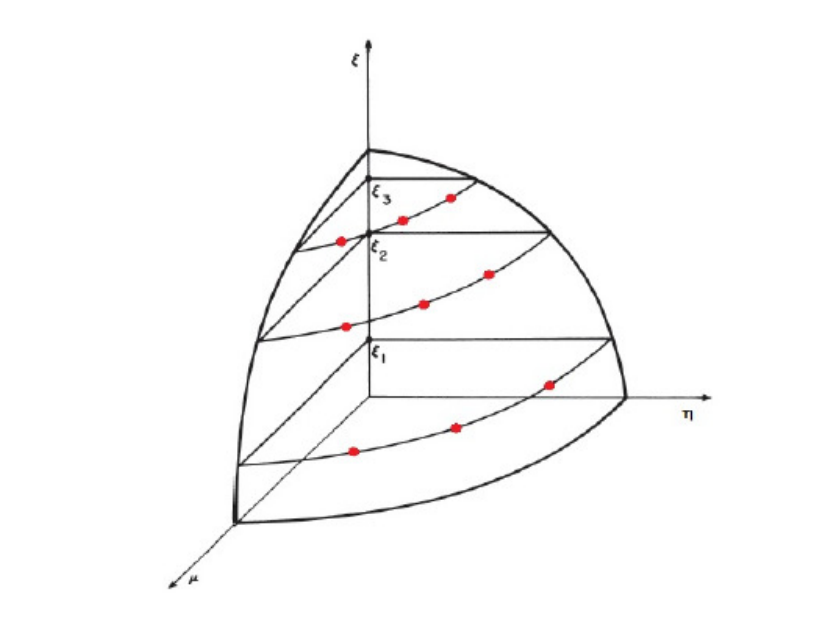}
\caption{Legendre-Chebyshev Quadrangular Scheme}
\label{pntn}
\end{figure}

The Legendre-Chebyshev quadrature schemes use a one-dimensional 
Gaussian quad\-ra\-ture to treat the polar variable $\theta$ and Chebyshev quadrature for the azimuthal
variable $\phi$ \cite{pleg-cheb, cacuci}.
Thus, in a three-dimensional geometry, the ordinates 
 $\xi=\cos\theta$ are given by the roots of the 
Legendre polynomials ($P_N$).
The azimuthal angular variable in each level is defined by the roots of the Chebyshev polynomials ($T_N$) of the first kind.
The Legendre-Chebyshev quadrangular scheme ($P_NT_N$) \cite{cacuci}
 defines  
$M=(N^{2}/4)$ discrete directions per octant, 
where for each level $\xi_i$ the corresponding directions have the same weight (Fig.~\ref{pntn}). 
 For the two-dimensional case, one coordinate is disregarded.
Still, the corresponding quadrature weights are given in terms of $w_i$, the weights of the Gauss-Legendre quadrature. 
For  the  Legendre-Chebyshev 
Triangular scheme ($P_NT_NS_N$) \cite{pleg-cheb}, on the other hand,  the 
$M=N(N+2)/8$ discrete directions per octant follow a pattern
 analogous  to the Level Symmetric quadrature scheme. 
The set of azimuthal angles for each level $\xi_i$ is defined as
\begin{equation}
\phi_{i,j}=\dfrac{\pi}{2}\left(1-\dfrac{N-2j-2i+3}{N-2i+2}\right),
\label{eq11-2016}
\end{equation}
such that, 
with weights  
\begin{equation}
w_{i,j}=\dfrac{w_i}{N-2i+2},
\label{eq13-2016}
\end{equation}
where $i=1,\dots,N/2$ and $j=1,\dots,N-2i+2$.

The  Quadruple Range quadrature ($QR$) scheme, proposed by Abu-Shumays \cite{abu1}, is a product quadrature. 
 As a general description, the polar angle $\theta$ is approximated by Gauss-Christoffel quadrature
 and the azimuthal direction $\phi\in[0,\pi/2]$ is defined as symmetric with respect to $\phi=\pi/4$. 
The coordinates $\phi_i$ and weights $w_i$ are found by the solution of a nonlinear system of equations, which is solved by 
Newton iterations \cite{abu1}  and
 the number of
 discrete ordinates per octant is $N_{\theta} \times N_{\phi}/2$, where
  $N_{\theta}$ is the order of the polar quadrature and  
 $N_{\phi}$ order of the azimuthal quadrature.
For a fixed order of quadrature, twice as many coordinates and weights are needed for the $\phi$ variable than for the $\theta$ variable, otherwise the quadrature would not be consistent \cite{abu1}.

Using the above mentioned schemes we \cite{jctt2016, anderson,physor}  investigated the property of 
 exactly  integrating the  highest order (direction cosines) polynomials. 
Such property is related to the 
 error in the numerical approximation of the integral term, in Eq.~(\ref{eq1}). 
As a result, it was found that the square Legendre-Chebyshev quadrature ($P_NT_N$) showed a better performance on this aspect, in the sense of integrating exactly higher-order polynomials for a fixed quadrature order $N$. It is a very relevant issue when analyzing the convergence of the discrete ordinate solution according to Madsen's theorem \cite{Madsen1971}, as we mentioned earlier in this text. On the other hand, the $(QR)$ quadrature scheme has shown better performance \cite{jctt2016} to mitigate the so-called ``ray effect'' - a well-known numerical issue associated with the discrete-ordinates method in multidimensional geometries. Later, Spence \cite{spence2015} proposed new approaches to generate ($QR$) schemes avoiding the ill-conditioning appearing in the previous formulation \cite{abu1}.
 In \cite{josadaque2020}, a complete study and discussion on the generation of the (QR) schemes was developed, through the use of non-classical orthogonal polynomials.

The influence of the quadrature schemes on the solution of the transport/RTE equation has been investigated in several of our works
\cite{mc2015,physor2018,lbb-kr-rdc2022}. It is important to note that, while using the $LQ_N$ schemes, we were able to consider only up to $36$ discrete directions per octant in our simulations. The alternative schemes allowed us to consider, for example, 
up to $98$ \cite{lbb-kr-rdc2022}, and $253$ \cite{nse-exact}. In fact, as the anisotropy of the medium increases, as in the case of optical tomography applications, we found it mandatory to increase the number of directions.

\subsection{The ADO-Nodal Formulation}
\label{adonodal}

We write the two-dimensional discrete ordinates approximation of Eq.~(\ref{eq1}) in Cartesian geometry as \cite{nse-exact,lbb-kr-rdc2022}
    \begin{equation}
\mu_{i} \frac{\partial }{\partial x}I(x,y,\Omega _{i})+ \eta_{i}\frac{\partial }{\partial y}I(x,y,\Omega_{i})+\beta I(x,y,\Omega_{i})=
 \frac{\sigma _{s}}{4\pi } \sum_{n=1}^{M}w_{n}I(x,y,\Omega _{n})p(\Omega_{n}\cdot \Omega_{i})+S(x,y),
\label{final}
    \end{equation}
for $i=1,...,M$, where $M$ is the total number of discrete directions 
in the upper four octants of the unit sphere and $w_{n}$  
are the associated weights, according to a chosen quadrature scheme;
$\beta=\kappa+\sigma_{s}$ is the extinction coefficient,  $\kappa$ 
is the absorption coefficient and  $\sigma_{s}$ is the scattering coefficient; 
$S(x,y)$ is the source term; $p (\Omega_{n} \cdot \Omega_{i})$ 
is the phase function of scattering from the incoming direction $\Omega_{i}$ to the scattered direction $\Omega_{n}$. We note that at this point, we are dealing with a system of first-order partial differential equations.

\begin{figure}[!ht]
    \centering
    \includegraphics[width=5.0cm]{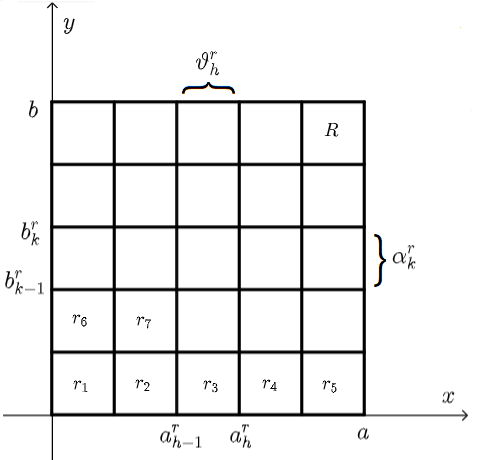}
    \caption{Domain subdivided into a mesh of HxK rectangular regions}
    \label{divided}
\end{figure}

We use a nodal method \cite{Badru1985} to solve Eq.~(\ref{final}), for each direction $i$, in the physical domain in the $(x,y)$ plane.  
 To do this, the domain is divided into $r=1, \cdots, R$ regions (nodes),
 where each region $r$ is defined for $(x,y) \in [a_{h-1},a_{h}]
 \times [b_{k-1},b_{k}]$ (see Fig.~\ref{divided}).

Classical nodal methods are coarse mesh methods based on the concept of transverse integration of one spatial variable at a time, such that the systems of partial differential equations arising from the discretization of the angular variable in the original problem are reduced to (coupled) systems of ordinary differential equations for each one of the spatial variables.

Since we use the ADO formulation developed in Section \ref{sec:ado} to solve the ODE system resulting from the application of the nodal scheme, we refer to this approach as the ADO-Nodal formulation \cite{lbb-pic-rdc2017}. 

\begin{figure}[!ht]
\centering
\includegraphics[scale=0.5]{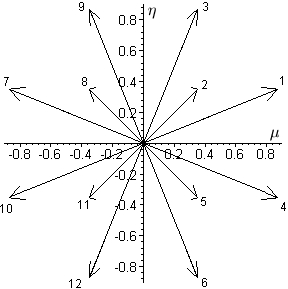}
\caption{Ordering Discrete Directions, $LQ_{4}$, $M=12$}
\label{xscheme}
\end{figure}

When deriving the two-dimensional model, we assume that intensities are the same at $\boldsymbol{\Omega_i}=(\xi_i, \mu_i, \eta_i)$ and $\boldsymbol{\Omega_i}=(-\xi_i, \mu_i, \eta_i)$. It will be used to deal with the phase-function for\-mu\-la\-tion \cite{nse-exact}. In addition to that, when working with the angular coordinates on the plane, to obtain, for instance, the average angular intensity along the $y$ direction, we follow \cite{lbb-lc-jf2011} and start ordering the discrete directions (for $i=1,\ldots, M/2$) such that (Fig.~\ref{xscheme})

\begin{eqnarray}
{\pmb \Omega}_{i} \!\!\!&=&\!\!\! (\mu_{i},\eta_{i})\\
{\pmb \Omega}_{i+M/2} \!\!\!&=&\!\!\! (-\mu_{i},\eta_{i}).
\end{eqnarray}

We refer to this ordering as $x$-scheme. Keeping all of that in mind,
we then let $\alpha^{r}_{k}=\left(b^{r}_{k}-b^{r}_{k-1}\right),$
multiply  Eq.~$(\ref{final})$ by $1/ \alpha^{r}_{k}$ and integrate over $y\in \left[b^{r}_{k-1},b^{r}_{k}\right]$
to get, for $i=1,..., M/2$

\begin{multline}
    \mu_{i} \frac{d}{d x} I_{yr}(x, \boldsymbol{\Omega}_{i}) + \beta_{r} I_{yr}(x, \boldsymbol{\Omega}_{i}) = \frac{\sigma_{sr}}{4\pi} \sum_{n=1}^{M/2} w_{n} \Big\{ I_{yr}\left(x, \boldsymbol{\Omega}_{n}\right)  \big[ p\left(\boldsymbol{\Omega}_{n} \cdot \boldsymbol{\Omega}_{i}\right) + p\left(\boldsymbol{\Omega}_{M+n} \cdot \boldsymbol{\Omega}_{i}\right) \big] \\
    + I_{yr}\left(x, \boldsymbol{\Omega}_{n+\frac{M}{2}}\right) \big[ p\left(\boldsymbol{\Omega}_{n+\frac{M}{2]}} \cdot \boldsymbol{\Omega}_{i}\right) + p\left(\boldsymbol{\Omega}_{M+n+\frac{M}{2}} \cdot \boldsymbol{\Omega}_{i}\right) \big] \Big\}  + Q_{y}(x, \boldsymbol{\Omega}_{i}),
  \label{y.c}
\end{multline}

\begin{multline}
    -\mu_{i} \frac{d}{d x} I_{yr}\left(x, \boldsymbol{\Omega}_{i+\frac{M}{2}}\right)  + \beta_{r} I_{yr}\left(x, \boldsymbol{\Omega}_{i+\frac{M}{2}}\right)\\
    = \frac{\sigma_{sr}}{4\pi} \sum_{n=1}^{M/2} w_{n} \Big\{ I_{yr}\left(x,
\boldsymbol{\Omega}_{n}\right)
    \big[ p\left(\boldsymbol{\Omega}_{n} \cdot \boldsymbol{\Omega}_{i+\frac{M}{2}}\right) + p\left(\boldsymbol{\Omega}_{M+n} \cdot \boldsymbol{\Omega}_{i+\frac{M}{2}}\right) \big]
    \\ + I_{yr}\left(x, \boldsymbol{\Omega}_{n+\frac{M}{2}}\right) \big[ p\left(\boldsymbol{\Omega}_{n+\frac{M}{2}} \cdot \boldsymbol{\Omega}_{i+\frac{M}{2}}\right) + p\left(\boldsymbol{\Omega}_{M+n+\frac{M}{2}} \cdot \boldsymbol{\Omega}_{i+\frac{M}{2}}\right) \big] \Big\} + Q_{y}(x, \boldsymbol{\Omega}_{i+\frac{M}{2}}).
    \label{y.d}
\end{multline}

Here, the average intensity along the variable $y$ is
\begin{equation}
I_{yr}(x,\boldsymbol{\Omega}_{i})=
\frac{1}{\alpha _{k}^{r}}\int_{b_{k-1}^{r}}^{b_{k}^{r}}
I_{r}(x,y,\boldsymbol{\Omega}_{i})dy
\end{equation}
and
\begin{equation}
Q_{yr}(x,\boldsymbol{\Omega}_{i}) =  S_{yr}(x)
- \frac{\eta _{i}}{\alpha_{k}^{r}}\left[I_{r}\left(x,b_{k}^{r},
{\boldsymbol{\Omega}}_{i}\right)
-I_{r}\left(x,b_{k-1}^{r},{\boldsymbol{\Omega}}_{i}\right)\right].
\label{unknwon}
\end{equation}

It is worth noting, regarding the integration performed on the variable $y$, that the terms $I_{r}\left(x,b_{k-1}^{r},\boldsymbol{\Omega}_{i}\right)$ and 
$I_{r}\left(x, b_{k}^{r},\boldsymbol{\Omega}_{i}\right)$ %$I_{r}\left(a_{h-1}^{r},y,\boldsymbol{\Omega}_{i}\right)$ and
%$I_{r}\left(a_{h}^{r},y,\boldsymbol{\Omega}_{i}\right)$,
present in Eqs.~(\ref{unknwon}), refer to the intensities on the boundaries of the regions. These terms may be known only in the incoming directions on the boundaries $y=0$ and $y=b$. In the existing directions and at the boundaries of the internal nodes, these terms remain unknown. Thus, the resulting systems of equations for the average intensities have more unknowns than equations, as usual in nodal schemes. It requires the use of auxiliary equations to close the system. In this work, the unknown intensities are approximated by constant functions. We have investigated this issue and proposed alternative approximations
 \cite{Prolo2014, cromianski2019study} for the unknown (``leakage'') terms. It turns out the constant functions seem to be the best choice when balancing accuracy and computational cost.

We proceed similarly to derive the system for the average intensity along variable $x$. For more details, see \cite{lbb-kr-rdc2022}. It also requires a specific ordering for the discrete directions, we refer to as $y$-scheme, where, for $i=1,\ldots, M/2$,
\begin{eqnarray}
{\pmb \Omega}_{i} \!\!\!&=&\!\!\! (\mu_{i},\eta_{i})\\
{\pmb \Omega}_{i+M/2} \!\!\!&=&\!\!\! (\mu_{i},-\eta_{i}).
\end{eqnarray}
We end up with a coupled system (directions $x$ and $y$) of first-order ordinary differential equations. 

To develop the ADO formulation for each one of the ODE's equations, for $r=1,...,R$, and $i=1,...,M$, we seek for exponential solutions to the homogeneous version of Eqs.~(\ref{y.c}) and (\ref{y.d}) (and respectively for $I^{H}_{xr}(y,\boldsymbol{\Omega} _{i})$) as 
\begin{equation}
I^{H}_{yr}(x,\boldsymbol{\Omega} _{i}) = \Phi_{yr}(\nu_{r},\boldsymbol{\Omega}_{i})e^{\frac{-x}{\nu _{r}}}.
\label{sol_exponencial}
\end{equation}
Continuing, we define
\begin{equation}
    {U_{yr}} (\nu_{jr}, \boldsymbol{\Omega}_{i})=\Phi_{yr}\left(\nu_{jr}, \boldsymbol{\Omega}_{i}\right) + \Phi_{yr}\left(\nu_{jr}, \boldsymbol{\Omega}_{i+M/2}\right)
\end{equation}
and
\begin{equation}
    {V_{yr}} (\nu_{jr}, \boldsymbol{\Omega}_{i})=\Phi_{yr}\left(\nu_{jr}, \boldsymbol{\Omega}_{i}\right) - \Phi_{yr}\left(\nu_{jr}, \boldsymbol{\Omega}_{i+M/2}\right),
\end{equation}
and after many algebraic manipulations and extensive derivations \cite{lbb-pic-rdc2017, lbb-kr-rdc2022, nse-exact} we obtain an
eigenvalue problem 
\begin{equation}
    \frac{1}{\nu^{2}}U_{yr}(\nu_{j})= \left[ \mathbf{A}_{yr} \mathbf{B}_{yr}\right] U_{yr}(\nu_{j})
\end{equation}
where
\begin{equation}
    \mathbf{A}_{yr}(i,j) = \begin{cases}
        \frac{\sigma_{sr}}{4\pi \mu_{i}} w_{j} \bigg[ p\left(\boldsymbol{\Omega}_{j} \cdot \boldsymbol{\Omega}_{i}\right) - p\left(\boldsymbol{\Omega}_{j} \cdot \boldsymbol{\Omega}_{i+M/2}\right) \\
        \quad + p\left(\boldsymbol{\Omega}_{M+j} \cdot \boldsymbol{\Omega}_{i}\right) - p\left(\boldsymbol{\Omega}_{M+j} \cdot \boldsymbol{\Omega}_{i+M/2}\right) \bigg] - \frac{\beta_r}{\mu_{i}} & \text{if } i = j, \\
        \frac{\sigma_{sr}}{4\pi \mu_{i}} w_{j} \bigg[ p\left(\boldsymbol{\Omega}_{j} \cdot \boldsymbol{\Omega}_{i}\right) - p\left(\boldsymbol{\Omega}_{j} \cdot \boldsymbol{\Omega}_{i+M/2}\right) \\
        \quad + p\left(\boldsymbol{\Omega}_{M+j} \cdot \boldsymbol{\Omega}_{i}\right) - p\left(\boldsymbol{\Omega}_{M+j} \cdot \boldsymbol{\Omega}_{i+M/2}\right) \bigg] & \text{if } i \neq j.
    \end{cases}
\end{equation}
and
\begin{equation}
    \mathbf{B}_{yr}(i,j) = \begin{cases}
        \frac{\sigma_{sr}}{4\pi \mu_{i}} w_{j} \bigg[ p\left(\boldsymbol{\Omega}_{j} \cdot \boldsymbol{\Omega}_{i}\right) + p\left(\boldsymbol{\Omega}_{j} \cdot \boldsymbol{\Omega}_{i+M/2}\right) \\
        \quad + p\left(\boldsymbol{\Omega}_{M+j} \cdot \boldsymbol{\Omega}_{i}\right) + p\left(\boldsymbol{\Omega}_{M+j} \cdot \boldsymbol{\Omega}_{i+M/2}\right) \bigg] - \frac{\beta_r}{\mu_{i}} & \text{if } i = j, \\
        \frac{\sigma_{sr}}{4\pi \mu_{i}} w_{j} \bigg[ p\left(\boldsymbol{\Omega}_{j} \cdot \boldsymbol{\Omega}_{i}\right) + p\left(\boldsymbol{\Omega}_{j} \cdot \boldsymbol{\Omega}_{i+M/2}\right) \\
        \quad + p\left(\boldsymbol{\Omega}_{M+j} \cdot \boldsymbol{\Omega}_{i}\right) + p\left(\boldsymbol{\Omega}_{M+j} \cdot \boldsymbol{\Omega}_{i+M/2}\right) \bigg] & \text{if } i \neq j.
    \end{cases}
\end{equation}

The ADO method is a spectral method. The derivation of the eigenvalue problem is a key issue in this formulation. We note that with this derivation, the eigenvalue problem derived for the ADO-Nodal formulation for two-dimensional problems is quite general. It is written in terms of the exact representation of the phase-function. Once one changes the transport theory field and the exact representation of the scattering pattern changes, the formulation is already developed. We note that in previous works \cite{jctt2020,lbb-kr-rdc2022} the eigenvalue problem was derived based on the expansion of the phase-function for an arbitrary order of anisotropy $L$. In such case,
\begin{equation}
    A_{yr}(i,j) =
    \begin{cases}
        \displaystyle \frac{\sigma_{sr}}{2\pi \mu_{i}}
        \sum\limits_{l=0}^{L}
        \sum\limits_{\substack{p=0 \\ (l+p) \text{ par}}}^{l}
        (2-\delta_{0,p}) C^{p}_{l} P^{p}_{l}(\xi_{i}) w_{j} P^{p}_{l}(\xi_{j}) \zeta^{p}_{yl}(\Omega_{j})
        - \frac{\beta_{r}}{\mu_{i}},
        & \text{if} i = j, \\[12pt]
        \displaystyle \frac{\sigma_{sr}}{2\pi \mu_{i}}
        \sum\limits_{l=0}^{L}
        \sum\limits_{\substack{p=0 \\ (l+p) \text{ par}}}^{l}
        (2-\delta_{0,p}) C^{p}_{l} P^{p}_{l}(\xi_{i}) w_{j} P^{p}_{l}(\xi_{j}) \zeta^{p}_{yl}(\Omega_{j}),
        & \text{otherwise},
    \end{cases}
\end{equation}
with
\begin{equation}
    \zeta^{p}_{yl}(\Omega_{j})=\begin{cases}
 \textup{sen}(p\varphi _{j})\textup{sen}(p\varphi _{i})& \text{if p  even }  \\
\textup{cos}(p\varphi _{j})\textup{cos}(p\varphi _{i}) & \text{if p  odd }  
\end{cases}
\end{equation}
and
\begin{equation}
    B_{yr}(i,j) =
    \begin{cases}
        \displaystyle \frac{\sigma_{sr}}{2\pi \mu_{i}}
        \sum\limits_{l=0}^{L}
        \sum\limits_{\substack{p=0 \\ (l+p) \text{ par}}}^{l}
        (2-\delta_{0,p}) C^{p}_{l} P^{p}_{l}(\xi_{i}) w_{j} P^{p}_{l}(\xi_{j}) \Gamma^{p}_{yl}(\Omega_{j})
        - \frac{\beta_{r}}{\mu_{i}},
        & \text{if} i = j, \\[12pt]
        \displaystyle \frac{\sigma_{sr}}{2\pi \mu_{i}}
        \sum\limits_{l=0}^{L}
        \sum\limits_{\substack{p=0 \\ (l+p) \text{ par}}}^{l}
        (2-\delta_{0,p}) C^{p}_{l} P^{p}_{l}(\xi_{i}) w_{j} P^{p}_{l}(\xi_{j}) \Gamma^{p}_{yl}(\Omega_{j}),
        & \text{otherwise},
    \end{cases}
\end{equation}
where
\begin{equation}
    \Gamma ^{p}_{yl}(\Omega_{j})=\begin{cases}
 \textup{cos}(p\varphi _{j})\textup{cos}(p\varphi _{i})& \text{ if p even }  \\
\textup{sen}(p\varphi _{j})\textup{sen}(p\varphi _{i}) & \text{if p  odd }
\end{cases}
\end{equation}
$i=1,...,M/2$.

In this subsection, through the two-dimensional transport equation, we generated a system of one-dimensional equations, allowing the use of
 ADO method \cite{ado} for its solution.
 The homogeneous solution obtained by this method is constructed in terms of the solution of an eigensystem.
It is relevant to stress that the ADO method,  when combined with the multidimensional quadratures studied, preserves the property of reduction of the order of the associated eigenvalue problems, thereby providing solutions with a higher degree of computational efficiency.

We can write the homogeneous solution for the average intensity along the direction $y$, region (node) $r$ in the form
\begin{equation}
    I_{yr}^{H}\left( x, \boldsymbol{\Omega}_{i} \right) = \sum_{j=1}^{M/2} \bigg[ A_{j,r} \, \Phi_{yr}\left( \nu_{jr}, \boldsymbol{\Omega}_{i} \right) e^{-\frac{(x - a_{h-1}^{r})}{\nu_{jr}}}
    + A_{j+\frac{M}{2},r} \, \Phi_{yr}\left( \nu_{jr}, \boldsymbol{\Omega}_{i+\frac{M}{2}} \right) e^{-\frac{(a_{h}^{r} - x)}{\nu_{jr}}} \bigg],
    \label{homogeneo_Y_1}
\end{equation}
\begin{equation}
    I_{yr}^{H}\left( x, \boldsymbol{\Omega}_{i+\frac{M}{2}} \right) = \sum_{j=1}^{M/2} \bigg[ A_{j,r} \, \Phi_{jr}\left( \nu_{jr}, \boldsymbol{\Omega}_{i+\frac{M}{2}} \right) e^{-\frac{(x - a_{h-1}^{r})}{\nu_{jr}}}
    + A_{j+\frac{M}{2},r} \, \Phi_{yr}\left( \nu_{jr}, \boldsymbol{\Omega}_{i} \right) e^{-\frac{(a_{h}^{r} - x)}{\nu_{jr}}} \bigg],
    \label{homogeneo_Y_2}
\end{equation}
for $i=1,\cdots, M/2$, $(x,y) \in [a_{h-1}^{r},a_{h}^{r}] \times
[b_{k-1}^{r},b_{b}^{r}]$. 
The arbitrary coefficients $A_{j,r}$ and $A_{j+\frac{M}{2},r}$ 
are to be determined (as well as the superposition coefficients of the homogeneous solution for 
$I_{xr}^{H}\left( y, \boldsymbol{\Omega}_{i} \right)$)
from the solution of a linear system.

%\begin{equation}
%    \phi_{yr}(x)=\frac{1}{4}\sum_{n=1}^{M/2}w_{n}\left[I_{yr}(x, %\Omega_{n})+I_{yr}(x, \Omega_{n+\frac{M}{2}})\right].
%\end{equation}

%Analogously for $I_{xr}(y)$ and $\phi_{xr}(y)$.

%The Henyey-Greenstein scattering function:

%\begin{equation}
%    p (\Omega_{n} \cdot \Omega_{i})= \frac{1-g^{2}}{\left[1 +g^{2} -2g %(\Omega_{n} \cdot \Omega_{i}) \right]^{\frac{3}{2}}}, 
%    \label{HG_exata1}
%\end{equation}

The general solution to the problem is obtained by adding the homogeneous solution to the particular solution. To propose the particular solution, it is first necessary to define the source terms as in Eqs. (\ref{y.c}), (\ref{y.d}), which include unknown intensities. Defining these components requires the use of auxiliary equations. According to previous works \cite{lbb-pic-rdc2017, cromianski2019study, lbb-kr-rdc2022}, a good choice is defining these auxiliary equations by assuming the unknown terms as constant functions, as we mention earlier in this text.

Thus, the unknown intensities in the $i$ directions in each region $r$,  are expressed by
\begin{equation}
    I_{r}(a_{h}^{r}, y, \boldsymbol{\Omega}_{i})=C_{h,r,i} , \quad
    I_{r}(a_{h-1}^{r}, y, \boldsymbol{\Omega}_{i}) =C_{h-1.r,i},
\end{equation}
and 
\begin{equation}
    I_{r}(x, b_{k}^{r}, \boldsymbol{\Omega}_{i})=D_{k,r,i} , \quad
    I_{r}(x, b_{k-1}^{r}, \boldsymbol{\Omega}_{i}) =D_{k-1.r,i},
\end{equation}
for $i=1,\ldots,M$ and $r=1,\ldots,R$.

Defining the unknown intensities at the boundaries of the regions as constants as well as $S_{yr}(x)$ in Eq.~(\ref{unknwon}), allows us the choice of constant approximations for the particular solution, that is,
\begin{equation}
    I_{yr}^{P}(x,\boldsymbol{\Omega}_{i}) =Z_{i,r},
    \label{particular_y}
\end{equation}
\begin{equation}
    I^{P}_{xr}(y,\boldsymbol{\Omega}_{i}) =W_{i,r}.
    \label{particular_x}
\end{equation}

Thus, we can write the general solution for the average intensity along the $y$ direction, by adding the homogeneous (Eqs. (\ref{homogeneo_Y_1}) and (\ref{homogeneo_Y_2})) and particular solution (Eq.~(\ref{particular_y})), taking into account the defined ordering of directions ($x$-scheme), for $i=1,\ldots, M/2$, as
\begin{multline}
    I_{yr}\left ( x,\boldsymbol{\Omega} _{i} \right )=\sum_{j=1}^{M/2}\left[ A_{j,r}\Phi _{yr}\left (\nu_{yr} ,\boldsymbol{\Omega} _{i}\right ) e^{-\left ( x-a_{h-1}^{r} \right )/\nu_{yr} }\right.+\\\left.A_{j+M/2,r}\Phi _{yr}\left ( \nu _{jr} ,\boldsymbol{\Omega} _{i+M/2}\right ) e^{-\left ( a_{h}^{r}-x \right) /\nu_{jr}}\right]+Z_{i,r}
       \label{geral_Y_1}
\end{multline}
and
\begin{multline}
    I_{yr}\left ( x,\boldsymbol{\Omega} _{i+M/2} \right )=\sum_{j=1}^{M/2}\left [ A_{j,r}\Phi _{yr}\left (\nu_{yr} ,\boldsymbol{\Omega} _{i+M/2}\right ) e^{-\left ( x-a_{h-1}^{r} \right )/\nu_{jr} }\right.+\\\left. A_{j+M/2,r}\Phi _{yr}\left ( \nu _{yr} ,\boldsymbol{\Omega} _{i}\right ) e^{-\left ( a_{h}^{r}-x \right )/\nu_{jr}}\right ]+ Z_{i+ M/2,r}
       \label{geral_Y_2}
\end{multline}
for $x\in[a_{h-1}^{r},a_{h}^{r}]$.

For developing the formulation for the average intensity in the $x$ direction, we follow a similar procedure \cite{lbb-kr-rdc2022} and express the general solution, assuming the previously defined ordering directions ($y$-scheme), in the form
\begin{multline}
    I_{xr}\left ( y,\boldsymbol{\Omega} _{i} \right )=\sum_{j=1}^{M/2}\left [ B_{j,r}\Phi _{xr}\left (\gamma_{jr} ,\boldsymbol{\Omega} _{i}\right ) e^{-\left ( y-b_{k-1}^{r} \right )/\gamma_{jr} }\right.\\+\left. B_{j+M/2,r}\Phi _{xr}\left ( \gamma _{jr} ,\boldsymbol{\Omega} _{i+M/2}\right ) e^{-\left ( b_{k}^{r}-y \right )/\gamma_{jr}}\right ] +W_{i,r}
    \label{geral_X_1}
\end{multline}
and
\begin{multline}
    I_{xr}\left ( y,\boldsymbol{\Omega} _{i+M/2} \right )=\sum_{j=1}^{M/2}\left [ B_{j,r}\Phi _{xr}\left (\gamma_{jr} ,\boldsymbol{\Omega} _{i+M/2}\right ) e^{-\left ( y-b_{k-1}^{r} \right )/\gamma_{jr} } +\right.\\\left. B_{j+M/2,r}\Phi _{xr}\left ( \gamma _{jr} ,\boldsymbol{\Omega} _{i}\right ) e^{-\left ( b_{k}^{r}-y \right )/\gamma_{jr}}\right ] + W_{i+M/2,r}
       \label{geral_X_2}
\end{multline}
for $i=1,\ldots, M/2$, $r=1,\ldots,R$, and $y\in[b_{k-1}^{r},b_{k}^{r}]$.
Still, $\gamma_{jr}$ are the associated eigenvalues.

To completely establish the average intensities in the variables $x$ and $y$ we need to solve a $4M(HK)$ linear system that results to be only for the arbitrary coefficients  ($A_j$, $B_j$, $Z_j$, and $W_j$, with $j=1,\ldots, M$). 
The order of the linear system is defined by the number of discrete directions as well as the number of the spatial nodes. High-quality solutions are expected as both, the number of discrete directions and
the refinement of the spatial mesh increase. We have investigated the performance of direct and iterative
methods, for the solution of such linear systems, along with domain decomposition techniques and parallel implementation \cite{mc2015, mec2021, coam}. Alternative arrangements in the configuration
of the equations allowed solutions to higher-order systems. A dependence on the type of
the quadrature scheme as well as the class of problems to be solved (neutron or radiation
problems, for instance) directly affects the final choice of the numerical algorithm \cite{coam, physor2018}.
Iterative methods such as GMRES, Loose GMRES, and TFQMR were successful in
neutron applications for different choices of quadrature schemes. For the thermal radiation problems \cite{lbb-kr-rdc2022} those methods were unable to satisfactorily solve the linear system. The
alternative was to obtain its solution via direct methods (LU factorization). Due to its
high sparsity, which leads to unacceptable levels of fill-in the factorization, a reordering of
the equations of the system was necessary to allow an application of a sparse block diagonal
subroutine \cite{mec2021, lbb-kr-rdc2022, coam}.

\section{Neutron Transport}
\label{neutrons}

An interesting aspect of BE is its applicability in different areas. Specific literature is usually found for each of these applications. Due to inherent features of the radiation interaction with the media, there are differences on the nomenclature and quantities of interest for each case. While the ADO-Nodal formulation was developed in the previous sections in the context of radiative transfer applications, we find interesting to stress particular aspects of the model and some issues we investigated in the context of the two-dimensional  neutron transport equation.

We begin with the time-independent neutron transport equation, which considers the distribution of the particles in non-multiplying homogeneous media, with one energy group, written as follows
\begin{eqnarray}
{\Omega \cdot \nabla \Psi({\bf r},{\boldsymbol \Omega})} + \sigma_t \Psi({\bf r},{\boldsymbol \Omega})= \int_{S}\sigma_s({\bf r},{\boldsymbol \Omega'} \cdot {\boldsymbol \Omega})\Psi({\bf r},{\boldsymbol \Omega'}) \,d{\boldsymbol \Omega'} + Q({\bf r},{\boldsymbol \Omega}),\label{eq1-neutron}
\end{eqnarray}
where now 
$\Psi({\bf r},{\boldsymbol \Omega})$ is the angular flux at ${\bf r}=(x,y,z)$ along direction 
${\boldsymbol \Omega}$; ${\boldsymbol \Omega}=(\mu,\eta,\xi)$ represents the direction of the particle as a vector on the unit sphere $S$;
$\sigma_t$ represents the total macroscopic cross section; 
$\sigma_s({\bf r},{\boldsymbol \Omega'} \cdot {\boldsymbol \Omega})$ is referred to as the differential scattering macroscopic cross section;
$Q({\bf r},{\boldsymbol \Omega})$ is the fixed neutron source term.
As for the RTE case discussed in Sec.~(\ref{sec:rte}), the equation will be supplemented with boundary conditions.

\begin{figure}[!ht]
\centering
\includegraphics[scale=0.4]{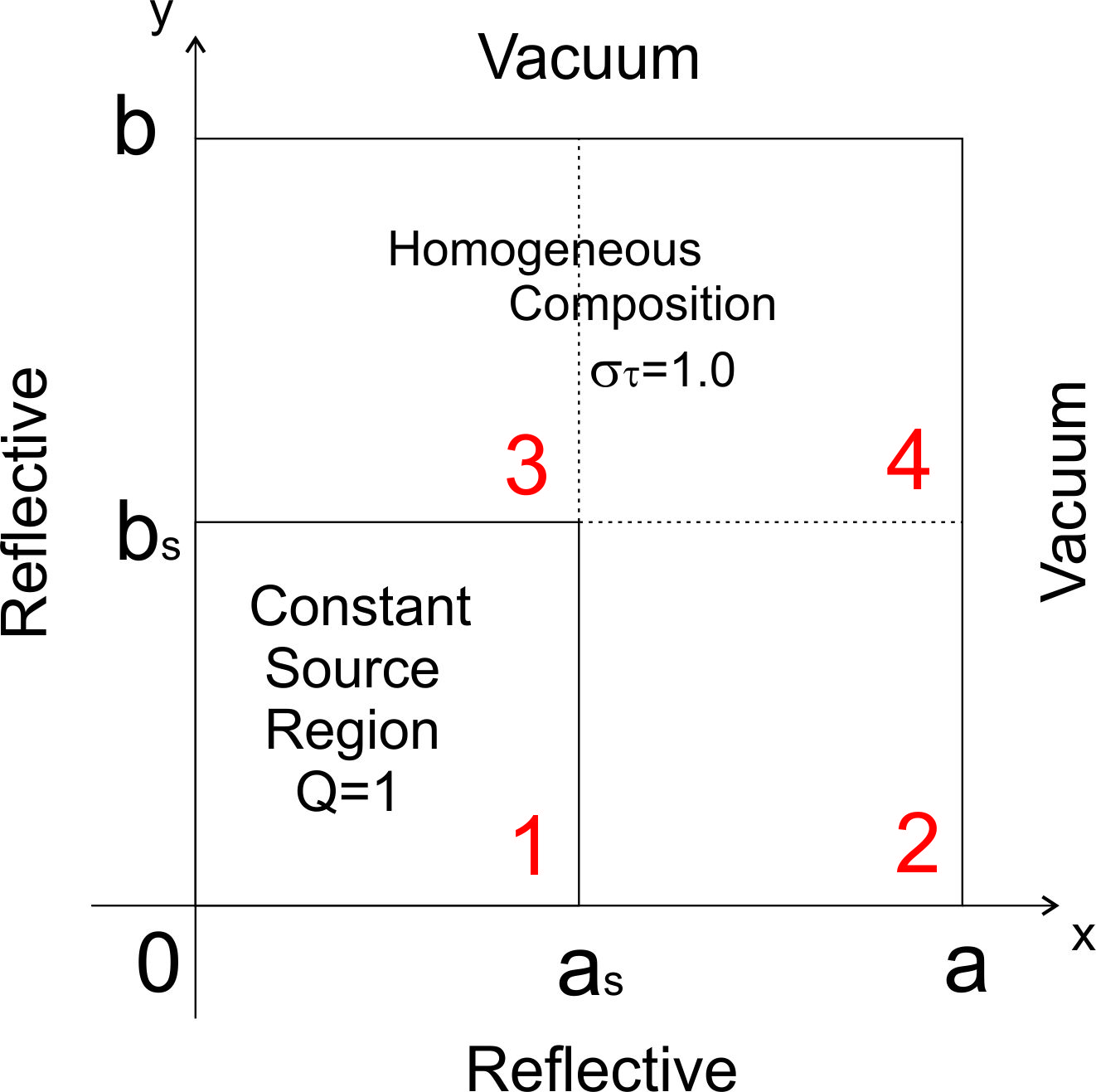}
\caption{Two-dimensional geometry.}
\label{bench}
\end{figure}

In fact, the first works on the application of the ADO method to two-dimensional geometry problems were devoted to neutron transport
\cite{lbb-lc-jf2011, Picoloto2015}. Much work has been done with respect to the solution of a known benchmark problem in reactor shielding applications. We describe the geometry of the problem in Fig.~\ref{bench}. Comparisons \cite{lbb-pic-rdc2017} with alternative nodal schemes available in the literature have shown a better performance of the ADO method in coarse meshes, as well as an expressive gain in computational time. 

However, the idea of bringing these applications to the attention here has to do with the analysis of the asymptotic error of the spatial discretization of the discrete-ordinates approximation of the two-dimensional neutron transport equation.

Madsen \cite{Madsen1971} discussed the pointwise convergence of the three-dimensional discrete ordinates method, showing, in particular, its relation with the truncation error in the scattering kernel approximation. We worked on this direction \cite{anderson,jctt2016,physor, josadaque2020}.

Using Richardson Extrapolation \cite{richardson}, a spatial asymptotic analysis was  made \cite{jctt2016}
 in order to determine the convergence order of nodal methods and a
spatial reference solution. 
Numerical results for region-averaged scalar flux were generated for comparisons between the ADO method and the AHOT-N0 \cite{ahot}, another nodal technique.

We investigated a model problem which
 consists of a homogeneous square domain with sides $a=b=1~ cm$,
with a source $Q(x,y)=1.0$ in the region $[0, 0.5]\times[0, 0.5]$,
as shown in Fig.~\ref{bench}.
The total cross section in the homogeneous domain is $\sigma_{t}=1.0~~cm^{-1}$
and we consider two different configurations where the scattering cross
section assumes two values:
$\sigma_{s}=0.9~~cm^{-1}$ in the first case and
 $\sigma_{s}=0.3~~cm^{-1}$ in the other.

\begin{figure}[!ht]
\centering
\includegraphics[scale=0.6]{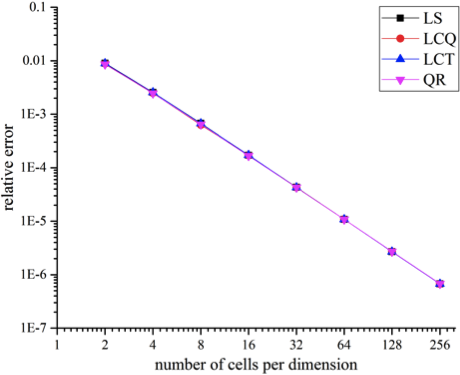}
\caption{Relative Error Computed Against Spatially-extrapolated Reference Solution (Region 1)}
\label{quadratic}
\end{figure}

\begin{figure}[!ht]
\centering
\includegraphics[scale=0.4]{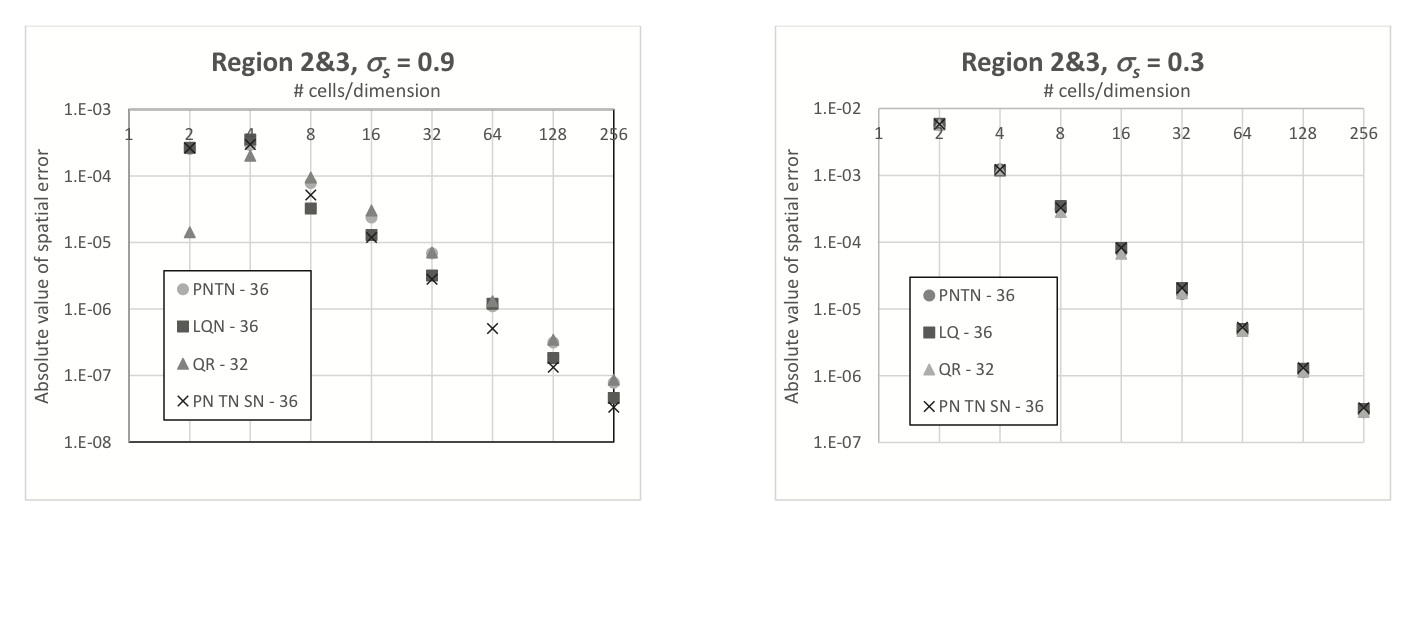}
\caption{Scalar flux absolute error against reference solution 
(Regions 2 and 3)}
\label{regiao3}
\end{figure}

\begin{figure}[!ht]
\begin{center}
\includegraphics[scale=0.4]{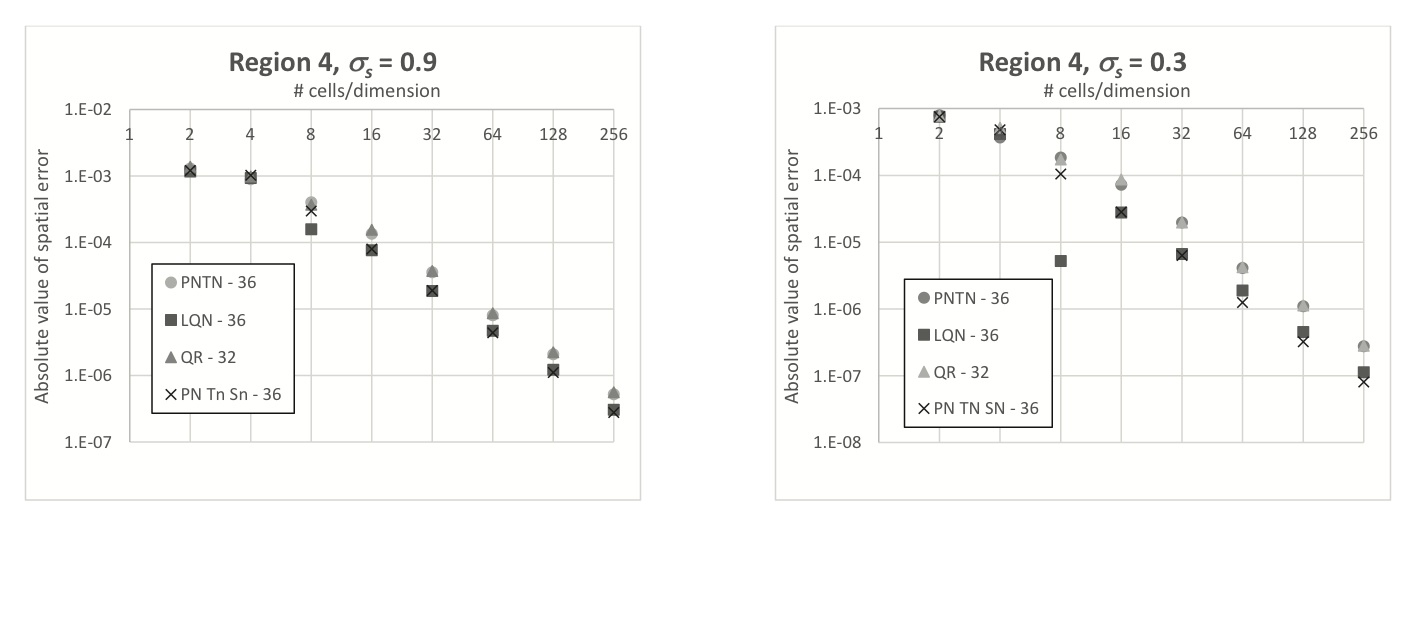}
\caption{Scalar flux absolute error against reference solution (Region 4)}
\label{regiao4}
\end{center}
\end{figure}

We consider each one of the four regions as a node when implementing
 the ADO method, and
to analyze the spatial convergence aspect we utilize
 $2^{j} \times 2^{j}$, $j= 1, \ldots, 8$, uniform mesh,
 when using the AHOT-N0 method. We investigate the spatial-discretization
 convergence
 through the application of the four quadrature schemes: $LQ_{N}$ \cite{lewis1984},
 $P_{N} T_{N} S_{N}$ \cite{pleg-cheb}, $P_{N} T_{N}$ \cite{cacuci} and
 QR \cite{abu1}  to the sequence of refined meshes in AHOT-N0. 
 The maximum number of directions per octant (ND) considered
 was $105$.
 Details are given in \cite{jctt2016}, including tables listing the numerical results. Here we reproduce some results given in that work, in Figs.~\ref{quadratic}, 
\ref{regiao3}, and \ref{regiao4}.

We proceeded \cite{jctt2016} with a spatial asymptotic analysis
 in order to determine the convergence order $p$  and a
spatial reference solution. For this purpose,
 we use the AHOT-N0 \cite{ahot} results and Richardson extrapolation \cite{richardson}
 to estimate an asymptotically extrapolated reference solution.
For a given number of directions per octant, ND, we assume
 $\phi(h)$ as a numerical approximation of order $p$ to
 the extrapolated (reference) solution, $\phi_{ref}$(ND). In this way,
 we write
\begin{eqnarray}
\phi(h)=\phi_{ref}+\alpha_ph^p+O(h^{p+1}),
\label{eq1-richardson}
\end{eqnarray}
and take evaluations on $h$, $rh$ and $r^{2}h$, where $r < 1$
 is a fixed factor,
\begin{eqnarray}
\phi(h)= \phi_{ref}+h^p\alpha_p+O(h^{p+1}),
\label{eq2}
\end{eqnarray}
\begin{eqnarray}
\phi(rh)= \phi_{ref}+(rh)^p\alpha_p+O(h^{p+1}),
\label{eq3}
\end{eqnarray}
\begin{eqnarray}
\phi(r^2h)= \phi_{ref}+(r^2h)^p\alpha_p+O(h^{p+1}).
\label{eq7}
\end{eqnarray}
After some algebraic manipulation
we can estimate the spatial convergence order $p$ as
\begin{eqnarray}
p=\dfrac{\log\left(\dfrac{\phi(r^2h)-\phi(rh)}{\phi(rh)-\phi(h)}\right)}{\log r},
\label{eq11}
\end{eqnarray}
%\begin{eqnarray}
%(r)^p=\dfrac{\phi(h)-\phi_{ref}}{\phi(rh)-\phi_{ref}}
%\label{eq6}
%\end{eqnarray}
and the spatial reference solution as
\begin{eqnarray}
\phi_{ref}=\left( \dfrac{r^p \phi(rh)-\phi(h)}{r^p-1}\right).
\label{eq13}
\end{eqnarray}

Results shown as $p$ order, and the resulting reference solution were
 calculated from the three finest  meshes of data with all
 digits available.
We found that the absolute error computed against the
 spatially-extrapolated reference solution for AHOT-N0 results,
 for the source region, it exhibits two features:
(i) quadratic convergence with mesh refinement,
 (ii) that is practically independent of the quadrature choice.
 These two features are illustrated, for $\sigma_{s}=0.9~~cm^{-1}$, in Fig.~\ref{quadratic}.
 
 \newpage

 On the other hand, for Regions $2,3$ and $4$,
   a tighter convergence criterion
 of order $10^{-11}$
 for refined meshes ($128 \times 128$ and $256 \times 256$)
 were required to indicate quadratic convergence
 order.
 Still,
 the behavior for each quadrature
 is somewhat different (see Figs.~\ref{regiao3} and \ref{regiao4}).

  We found from the results that asymptotic convergence
 in a strict sense ($p$ approaching a
 constant value) is not clearly established. The reason for this behavior is not
 evident and has deserved further investigation, nevertheless the fact that for
 the vast majority of cases considered the computed $p \sim 2$ is satisfactory.

\subsection{Numerical Results and Contributions}

The ADO formulation, extended to the solution of two-dimensional problems through the use of nodal schemes, is referred to as ADO-Nodal and is distinguished from other nodal schemes in that it retains explicit expressions for the average intensities in terms of each spatial variable. In addition, and very significantly, it has been associated with a class of quadrature schemes (another notable feature), retaining the order of the eigenvalue problem at half the number of directions. In the derivation presented in 
Section \ref{adonodal}, the eigenvalue problem was extended to allow the formulation to be applied to problems with an arbitrary degree of anisotropy using both the expanded and exact formulation. It represents another feature, as far as we know, singular in comparison with nodal schemes available in the literature -- without forgetting the issue that scattering is a significant challenge in these models. Numerical results that confirm such analysis may be found in the works listed in this text.

%*************************************************************
%***************************************************************

\section{A Taste of the Linearized Boltzmann Equation}
\label{sec:lbe}

Internal rarefied gas flows define a field of major interest in the
general area of rarefied gas dynamics \cite{cercig1969, Cercig1988}.
We have seen an increased interest in this area because
of applications, for example, as related
to micro-electro-mechanical machines (MEMS), where the Boltzmann equation or a model equation is required in order to describe
well gas-flow and heat-flow mechanisms. 
In many cases, the flow conditions are in the transition regime, and as a
result, the well-known and commonly used Navier-Stokes equations can not
be applied. In these cases, the Boltzmann equation or suitable kinetic
models should be utilized.

The ADO 
method \cite{ado} has been shown \cite{unified, tjump, pof} to be a useful method for solving a class 
of basic problems in this field. For example, many of the classical problems 
based
on the BGK model \cite{bgk} have been solved in a unified and especially accurate 
way with the ADO method. Following the beginning work with the BGK model, 
we extended the ADO method in order to solve many of the basic problems in 
rarefied gas dynamics that were defined in terms of the variable collision 
frequency model (CLF model) of  Cercignani \cite{clf}  and Loyalka and Ferziger \cite{ferziger}. 
Having seen that the ADO method was a convenient computational method for 
solving
problems based on the CLF model and seeking improved results for physical
quantities of interest, we moved forward \cite{some} and  
defined explicitly two extended model equations (CES model and the CEBS model) based on approximated forms
of the LBE.  We made use of exact solutions of the 
homogeneous and of the inhomogeneous LBE to provide a systematic
basis for approximating
the scattering kernel in the LBE. 

 Although intensive numerical methods can be used, as can the Monte Carlo method, to solve practical problems based
on the LBE, for rigid-sphere collisions, our goal was to explore a class of model equations that can
be solved (essentially) analytically to yield results good enough for selected applications and for producing benchmark solutions. We provided a systematic derivation of kinetic models and report the
final forms in sufficient detail that researchers seeking to develop numerical 
algorithms to solve practical problems could have a clearly defined 
starting point.

%We note here that while we have found success
%with our ADO method, the kernel of the integral operator of the LBE is not
%continuous \cite{pekeris} in the relevant variables, and for this reason we avoid the
%use of a global quadrature method for evaluating integrals the integrand of
%which can be discontinuous at any point ($c=c'$) in the domain of %definition. 

In \cite{some}, following \cite{pekeris, hickey}, we consider the 
LBE written
as
\begin{equation}
c\mu\frac{\partial\,\,}{\partial x}h(x,\boldsymbol{c})=\sigma_0^2 n_0 \pi^{1/2}
{L}\{h\}(x,\boldsymbol{c}),
\label{1-ces}
\end{equation}
where $c(2kT_0/m)^{1/2}$ is the magnitude of a particle velocity, $x$
is the spatial variable (measured in $cm$) and $h(x,\boldsymbol{c})$
defines the perturbation from the equilibrium distribution 
\begin{equation}
f _0(c)=n_0\big[m/(2\pi kT_0)\big]^{3/2}\ope^{-c^2}.
\label{2-ces}
\end{equation}
Here $n_0$ is the (constant) density of gas particles, each of mass $m$, 
$k$ is the
Boltzmann constant and $T_0$ is a (constant) reference temperature.
We note that while $h(x,\boldsymbol{c})$ defines the focus
of our attention, the actual distribution function 
$f(x,\boldsymbol{c})$ in this formulation is expressed as 
\begin{equation}
f(x,\boldsymbol{c})=f _0(c)[1+h(x,\boldsymbol{c})].
\label{3-ces}
\end{equation}
In regard to Eq.~(\ref{1-ces}), we state that \cite{pekeris,hickey}
 $\sigma_0$ is the collision diameter of the 
gas particles (in the
rigid-sphere approximation) and that the collision process is described
by
\begin{equation}
{L}\{h\}(x,\boldsymbol{c})=-\nu(c)h(x,\boldsymbol{c})+
\int_0^\infty\int_{-1}^1 \int_0^{2\pi}\ope^{-{c'}^2} h(x,\boldsymbol{c}')
K(\boldsymbol{c}',\boldsymbol{c})
{c'}^2\opd\chi'
\opd\mu'\opd c',
\label{4-ces}
\end{equation}
where we use spherical coordinates ($c,\arccos \mu ,\chi$) to define
the (dimensionless) velocity vector, where
$\nu(c)$ is the ``collision frequency'' and where $ K(\boldsymbol{c}',\boldsymbol{c})$ 
is the scattering kernel.
It is clear that to proceed we require definitions of the 
scattering kernel $K(\boldsymbol{c}',\boldsymbol{c})$
and the collision frequency $\nu(c)$. In their work,
Pekeris and Alterman \cite{pekeris} used an expansion in
terms of Legendre polynomials (as functions of the scattering angle
between ``before'' and ``after'' directions) to describe the scattering
process. We use the 
spherical-harmonics addition theorem and write the 
scattering kernel as
\begin{equation}
K(\boldsymbol{c}',\boldsymbol{c})=\frac{1}{2\pi}\sum_{n=0}^\infty
\sum_{m=0}^n
\Big(\frac{2n+1}{2}\Big)
(2-\delta_{0,m})
P_n^m(\mu')
P_n^m(\mu)
k_n(c',c)\cos m(\chi'-\chi).
\label{5-ces}
\end{equation}
Here the {\it{normalized}} Legendre functions
\begin{equation}
P_n^m(\mu)=\left[\frac{(n-m)\,!}{(n+m)\,!}\right]^{1/2}{(1-\mu^2)}^{m/2}
\frac{\opd^m\,\,}{\opd\mu^m}P_n(\mu),\quad n\ge m,
\label{6-ces}
\end{equation}
where $P_n(\mu)$ denotes the usual Legendre polynomial,
are such that
\begin{equation}
\int_{-1}^1P_n^m(\mu)P_{n'}^m(\mu)\opd\mu=\Big(\frac{2}{2n+1}\Big)\delta_{n,n'}.
\label{7-ces}
\end{equation}
The basic elements $k_n(c',c)$ in the expansion of the scattering 
kernel are available from the paper of \cite{pekeris}
where explicit expressions are given for $n=$ 1 and 2.

We introduce
a mean-free path $l$ and
use the dimensionless variable $\tau=x/l$ to rewrite Eq.~(\ref{1-ces}) as
\begin{equation}
c\mu\frac{\partial\,\,}{\partial \tau}h(\tau,\boldsymbol{c})=\varepsilon
{L}\{h\}(\tau,\boldsymbol{c})
\label{11-ces}
\end{equation}
where the operator $L$ is defined by Eq.~(\ref{4-ces}) and
\begin{equation}
\varepsilon=\sigma_0^2 n_0 \pi^{1/2}l.
\label{12-ces}
\end{equation}
%While it might be convenient to have introduced the idea
%of a mean-free path, we must keep in mind that this quantity can not, at
%this point, be considered known since in reality it is a function of
%the actual solution we seek. Some workers choose to use a mean-free
%path based on viscosity for flow problems and a mean-free path based
%on thermal conductivity for heat-flow problems. In either case, the
%use of a appropriate mean-free
%path is especially important when working with model equations.
To have our final results in terms of the
real spatial variable $x$ (in $cm$), then what we use for a mean-free
path $l$ is not important if the molecular diameter $\sigma_0$ 
and the density $n_0$ are known. However, since these physical quantities
may not be known, some researchers choose to
work in terms of a specific mean-free path. We can mention two
convenient choices. 
For the Poiseuille-flow problem, we follow \cite{hickey} 
and found that \cite{some},
\begin{equation}
\varepsilon=\varepsilon_{\opp} = \frac{16}{15}\pi^{-1/2}
\int_0^\infty \ope^{-c^2}b(c)c^6\opd c.
\label{51}
\end{equation}
For the temperature-jump problem, following \cite{ferziger} we use
\begin{equation}
\varepsilon=\varepsilon_{\opt} = \frac{16}{15}\pi^{-1/2}
\int_0^\infty \ope^{-c^2}a(c)c^6\opd c.
\label{55}
\end{equation}
In regard to numerical work, we note that
Hermite cubic splines have been used \cite{chapman}
to solve the 
Chapman-Enskog integral equations for viscosity and heat conduction, making it possible to evaluate 
\begin{equation}
\varepsilon_{\opp} = 0.449027806...
\quad{\text{and}}\quad
\varepsilon_{\opt} = 0.679630049...\,.
\label{56-ces}
\end{equation}
Noting that the Prandtl number normally used in kinetic theory (in contrast
to the difference by 5/2 definition used in fluid dynamics) written as
\begin{equation}
\Pr=\frac{5}{2}\frac{\mu_* k}{m \lambda_*}
\label{57}
\end{equation}
can be expressed as
\begin{equation}
\Pr=\varepsilon_{\opp}/\varepsilon_{\opt},
\label{58-ces}
\end{equation}
and so using Eq.~(\ref{56-ces}) in Eq.~(\ref{58-ces}), we find the result
\begin{equation}
\Pr=0.660694457...
\label{59}
\end{equation}
which we believe to be correct (for the LBE and rigid-sphere collisions)
to all digits given.

\subsection{A Synthetic Kernel}

Even if we truncate the expansion of the scattering kernel given as Eq.~(\ref{5-ces})
after only a few terms, the problem of solving the resulting approximation of
the LBE is still difficult from a numerical point of view. The numerical
difficulty comes about basically because the components $k_n(c',c)$ 
required in Eq.~(\ref{5-ces}) have derivatives (even for small values of $n$) that are 
discontinuous at $c'=c.$ Keeping in mind that we intend
to implement our work numerically,  we seek to approximate the true kernel
by physically meaningful 
approximations that can be more easily incorporated into a numerical 
algorithm. 
The  variable collision frequency  (CLF) model, for instance, 
\cite{clf} and \cite{ferziger}, was used in \cite{tjump,pof} in  
order to try to
improve basic results available from the classical BGK model \cite{bgk}.
In \cite{some}, we extend the variant of the variable collision frequency model 
used in \cite{tjump, pof} 
by replacing the exact
components $k_n(c',c)$ by a truncated form of the scattering 
kernel $K(\boldsymbol{c}',\boldsymbol{c})$ 
with a more general synthetic approximation. To begin,
we truncate Eq.~(\ref{5-ces}) and write the synthetic scattering kernel as
\begin{equation}
F(\boldsymbol{c}',\boldsymbol{c})=\frac{1}{2\pi}\sum_{n=0}^N
\sum_{m=0}^n
\Big(\frac{2n+1}{2}\Big)
(2-\delta_{0,m})
P_n^m(\mu')
P_n^m(\mu)
f_n(c',c)\cos m(\chi'-\chi)
\label{27-ces}
\end{equation}
where, instead of using $k_n(c',c)$ as in Eq.~(\ref{5-ces}), we propose to use synthetic
approximations $f_n(c',c)$ which we express, initially for $N=2$, as 
\begin{equation}
f_0(c',c)=A_0(c')A_0(c)+B_0(c')B_0(c),
\label{28a}
\end{equation}
\begin{equation}
f_1(c',c)=A_1(c')A_1(c)+B_1(c')B_1(c)
\label{28b}
\end{equation}
and
\begin{equation}
f_2(c',c)=A_2(c')A_2(c),
\label{28c}
\end{equation}
where the functions $\{A_n(x),B_n(x)\}$ are to be determined. This way, we
now write our approximated balance equation as
\begin{equation}
c\mu\frac{\partial\,\,}{\partial \tau}h(\tau,\boldsymbol{c})=\varepsilon
{L^*}\{h\}(\tau,\boldsymbol{c})
\label{29}
\end{equation}
where
\begin{equation}
{L^*}\{h\}(\tau,\boldsymbol{c})=-\nu(c)h(\tau,\boldsymbol{c})+
\int_0^\infty\int_{-1}^1 \int_0^{2\pi} \ope^{-{c'}^2}h(\tau,\boldsymbol{c}')
F(\boldsymbol{c}',\boldsymbol{c})
{c'}^2\opd\chi'
\opd\mu'\opd c'.
\label{30}
\end{equation}
Here $F(\boldsymbol{c}',\boldsymbol{c})$ is given by Eq.~(\ref{27-ces}) with $N=2$. 

The definition of  the
components $f_n(c',c), n=0,1,2,$ of the $F(\boldsymbol{c}',\boldsymbol{c})$, detailed in \cite{some}, results from the requirement that exact solutions of the LBE satisfy also the model equation. 
This derivation involves the solution of a class of integral equations \cite{chapman}.  
Such formulation allowed us to propose two additional kinetic models, referred to as CES and CEBS, to the ones available in the literature by that time (BGK, S-model, CLF). 
 
In defining what we have called the CES model of the linearized Boltzmann
equation, we established five conditions on the synthetic 
kernel $F(\boldsymbol{c}',\boldsymbol{c})$ by insisting that the known solutions of the 
homogeneous LBE
be also solutions of the model equation. As we
have no more known solutions, we require other conditions to 
have a more general model equation. Since the classical problems of Poiseuille
flow and thermal-creep flow in a plane channel are normally defined in terms of an inhomogeneous version of the LBE. We  define our
new model equation by insisting that the particular solution required for
the LBE be also a particular solution of the inhomogeneous model equation. 
With that, we determined $f_n(c',c), n=0,1,2,3$, so as to define
a new synthetic kernel.  
The CEBS model has one
more term ($N=3$ rather than $N=2$) in the synthetic scattering kernel,
and it proved to be a good addition to the  
class of model equations already available for approximating the linearized
Boltzmann equation.

In summary, as we state in \cite{some}, in our work on the rarefied gas dynamics, we have made consistent use of the idea of approximating
the exact scattering kernel relevant to the linearized Boltzmann equation
for rigid-sphere interactions with a synthetic kernel that maintains
some basic properties of the exact kernel. To be clear, we insist 
that a model equation defined by an approximating synthetic kernel 
accept as solutions certain known solutions of the homogeneous LBE or
the inhomogeneous LBE relevant to forced flow in a plane channel. By
developing improved model equations 
(what we call the CES model and the CEBS model) 
that are amenable to analytical and simple numerical 
methods of solution, we provided conditions  that numerical results for 
applications of interest can be made available without the need of extensive computation
methods required when the LBE must be solved numerically. 
In addition to that, it allowed us to model and solve in a unified way 
a series of classical problems in RGD: using the same methodology, comparing the results of different kinetic models, providing concise and accurate solutions for such problems, including gas mixtures and the analysis of the Cercignani-Lampis boundary condition. Some of these results can be found in \cite{cabrera, siam, Scherer2009, Scherer2010, cesmodel}.

\subsection{Kinetic Models and the G-problem}

In solving a wide class of problems in the rarefied gas dynamics listed in Section \ref{sec:lbe}, algebraic manipulations led us to the application of the ADO method to the solution of an auxiliary problem we usually referred to as the ``G-problem'' that we write here in the form
\begin{equation}
\xi\frac{\partial}{\partial \tau}G(\tau,\xi)+G(\tau,\xi)=\int_{-\gamma}^\gamma
G(\tau,\xi')\,d\xi'\,,
\label{9-g}
\end{equation}
for $\xi \in (-\gamma,\gamma)$ and $\tau \in (0,\tau_0)$,
with boundary conditions, in general, written as 
\begin{equation}
G(0,\xi)=G_1(\xi)+G(0,-\xi)
\label{10a}
\end{equation}
and 
\begin{equation}
G(\tau_0,-\xi)=G_2(\xi)+G(\tau_0,\xi),
\label{10b}
\end{equation}
for $\xi \in (0,\gamma)$. Here, $\Psi(\xi)$ is called the characteristic function, which assumes different definitions depending on the problem. We note that for some applications \cite{ado, unified}, $\gamma$ is such that the integration interval is defined as $(-\infty, \infty)$.
The central aspect we want to point out is that the challenging solution of a 
``G-problem'' type was the motivation for proposing the ADO method \cite{ado}. The first paper with the derivation of the ADO method. Even more, 
the versatility of the ADO formulation in solving this more general type of problem (than the problem given by Eqs.~(\ref{eq:slab2}) and (\ref{rte_isr})) distinguishes the ADO method from other discrete ordinate methods usually associated only with the treatment of problems in the area of neutron transport and radiative transfer. It is worth to note that, the research on the kinetic models led to the solution of the LBE  
\cite{ceslbe, lbemodels} through the application of the ADO method to the scalar ``G problem'', Eq.~(\ref{9-g}), and its vector versions.

\section{Inverse Problems}
\label{sec:inverse}
The focus of this work was the development and application of solutions, using the ADO method, to solve direct problems. However, there are many inverse problems of interest in transport theory. For instance, we have investigated in recent years the reconstruction of radiation sources, which are fundamental to nuclear safety \cite{Pazinatto2020, Pazinatto2021}. Currently, we are studying the problem of reconstructing the absorption and scattering coefficients, which is essential to optical tomography \cite{aip2025}.

The primary benefit is that the method's analytical feature, which provides explicit expressions in terms of the spatial variables, significantly contributes to accuracy and computational time. These are fundamental factors in inverse problem techniques.

\section{Concluding Remarks and Ongoing Research}
\label{sec:concluding}
In this work, a brief overview of studies and advances in the solution of the discrete ordinates approximation of the linear Boltzmann equation in one and two spatial dimensions was presented. The focus was on the application areas of neutrons, photons, and rarefied gas dynamics. In particular, the successful applicability of the Analytical Discrete Ordinates (ADO) method to a series of particle transport problems was emphasized through the reference of selected  relevant publications. A broad literature review in the area is not within the scope of this work,
but rather to draw attention to research in this important area, conducted in Latin America and more specifically in Brazil.

Despite the advances and contributions we have sought to highlight, several important issues still deserve attention for the development of the method and the subject. In particular, domain decomposition methods are being investigated to address the large-scale linear system issue that is inherent to the formulation. Another area of study, still open, is the analysis of asymptotic error due to angular discretization — more specifically, the joint analysis of spatial-angular discretizations.
Finally, we intend to use the formulation with exact representation of the phase function in the problems of reconstruction of absorption and scattering parameters, of interest in optical tomography, treated by the ADO method, so far, only in one-dimensional geometry.

\end{document}